# Modeling functional resting-state brain networks through neural message passing on the human connectome

Julio A. Peraza-Goicolea[a], Eduardo Martínez-Montes[b,c*], Eduardo Aubert[b], Pedro A. Valdés-Hernández[d], and Roberto Mulet[a]

[a] *Group of Complex Systems and Statistical Physics, Department of Theoretical Physics, University of Havana, Havana, Cuba*

[b] *Neuroinformatics Department, Cuban Neuroscience Center, Havana, Cuba*

[c] *Advanced Center for Electrical and Electronic Engineering (AC3E), Universidad Técnica Federico Santa María, Valparaíso, Chile*

[d] *Neuronal Mass Dynamics Laboratory, Department of Biomedical Engineering, Florida International University, Miami, FL, USA*

***Corresponding author:**

Calle 190 #15202 entre Ave 25 y 27, Cubanacan, Playa, Habana 11600

Phone: +53 7 263 7252

E-mail: eduardo@cneuro.edu.cu (E. Martínez-Montes)

**Abstract**

Understanding the relationship between the structure and function of the human brain is one of the most important open questions in Neurosciences. In particular, Resting State Networks (RSN) and more specifically the Default Mode Network (DMN) of the brain, which are defined from the analysis of functional data lack a definitive justification consistent with the anatomical structure of the brain. In this work we show that a possible connection may naturally rest on the idea that information flows in the brain through a neural message-passing dynamics between macroscopic structures, like those defined by the human connectome (HC). In our model, each brain region in the HC is assumed to have a binary behavior (active or not), the strength of interactions among them is encoded in the anatomical connectivity matrix defined by the HC, and the dynamics of the system is defined by a neural message-passing algorithm, Belief Propagation (BP), working near the critical point of the human connectome. We show that in the absence of direct external stimuli the BP algorithm converges to a spatial map of activations that is similar to the DMN. Moreover, we computed, using Susceptibility Propagation (SP), the matrix of correlations between the different regions and show that the modules defined by a clustering of this matrix resemble several Resting States Networks determined experimentally. Both results suggest that the functional DMN and RSNs can be seen as simple consequences of the anatomical structure of the brain and a neural message-passing dynamics between macroscopic regions. We then show preliminary results indicating our predictions on how functional DMN maps change when the anatomical brain network suffers structural anomalies, like in Alzheimer's Disease and in lesions of the Corpus Callosum.

## 1. Introduction

Despite the efforts of the scientific community, the relationship between the structure and function of the brain is still an open question in Neuroscience. On one hand, the combination of diffusion spectrum imaging and tractography algorithms, allows to determine a macroscopic representation of the structural network of the brain (Hagmann et al., 2008). On the other hand, from functional magnetic resonance imaging (fMRI), it is possible to determine the resting-state functional connections between the brain regions at the macroscopic level (Michael D Greicius, Krasnow, Reiss, & Menon, 2003). However, consensus has not been reached about which is a plausible dynamical model of the brain function, able to quantitatively describe the relation between these structural and functional networks.

Recent studies have shown a recurrent finding of a set of brain regions that appear active in almost all brain states, which has been called the brain Default Mode Network (Raichle, 2015). Based on traditional experimental studies that related brain regions with different functions, several authors have explained the Default Mode Network (DMN) as a brain system that is preferentially active when individuals are not focused on external stimuli. The DMN has been claimed to be involved in processes with emotional support given activations of the Ventral Medial Prefrontal cortex (Simpson, Drevets, Snyder, Gusnard, & Raichle, 2001). Activations in the medial temporal lobe suggested the involvement in retrieval of memories from past experiences (Vincent et al. 2006; Buckner et al. 2008) while those in the Medial Prefrontal Cortex Dorsal suggested self-referring mental activity (Gusnard & Raichle, 2001). These two subsystems converged in the Posterior Cingulate Cortex which is a well-known hub for information integration in the brain (Buckner, Andrews-Hanna, & Schacter, 2008). In general, the evidence indicates that the activity in the DMN never shuts down, but it is just modulated during the resting-state (Raichle, 2015). Additionally, its functional elements can be differentially affected during the execution of a task by the nature of the action (for example, the presence or absence of emotional components) (Andrews-Hanna, Reidler, Sepulcre, Poulin, & Buckner, 2010; Gusnard & Raichle, 2001). However, it is still unclear the functional role (if any) of the DMN as a whole, or even why it is present in almost all brain states (Raichle, 2015).

The appearance of anatomical and functional brain networks brought to neuroscience the introduction of concepts and techniques from Network Theory and Complex Systems. For example, in the last decade many studies have shown that structural networks of the brain are characterized by high cluster index and modularity combined with a high efficiency and a short path length (Bullmore & Sporns, 2009; Rubinov & Sporns, 2010; Sporns, 2011). Similar results have also been found when studying functional networks (see e.g. Fallani et al., 2010; Papo et al., 2014; Sanabria-Diaz et al., 2013; Sporns, 2018, and references therein). However, most of these studies are descriptive in nature and theoretical explanations of the relationship between the properties of the networks and normal or pathological brain states are not straightforward.

As a first attempt to understand brain function from these perspectives, in 1996 Per Bak proposed that the complexity of brain function should be associated with the existence of a critical dynamics in a sense similar to that found in a second order phase transition (Bak, 1996). The first clear evidence of critical dynamics in the brain was given by the experiments of Beggs and Plenz in 2003 (Beggs & Plenz, 2003). They identified a mode of spontaneous activity in cortical networks from organotypic cultures and acute slices, called "neuronal avalanches". They found that this neural activity has numerous points of contact with Self-Organized Criticality theory. For example, the distribution of avalanche sizes, from all cortical networks, showed no characteristic scale and could be described by power laws, resembling those found near the critical point of a second order phase transition in systems such as sandpile avalanches, and magnetism. Years later, Dante Chialvo and colleagues showed that also fMRI data gathered in humans in the absence of external stimuli, exhibits these scale-free properties (Chialvo, 2010; Expert et al., 2011), supporting a picture consistent with a brain at rest being near a critical point (Tagliazucchi, Balenzuela, Fraiman, & Chialvo, 2012).

Second order phase transitions have been largely studied in Statistical Physics, where multiple computational models have emerged in the line of these previous empirical findings (Fraiman et al. 2009; Kitzbichler et al. 2009; Chialvo 2010; Deco and Jirsa 2012). In this respect, the Ising model, a well-known paradigm in Physics, has been used in the modeling of neural networks (Amit, Gutfreund, and Sompolinsky 1985; Amit 1989; Das et al. 2014), and more specifically to describe the macroscopic brain activity in brain networks at the appropriate critical point (Fraiman et al. 2009;Deco, Senden, and Jirsa 2012; Marinazzo et al. 2014). Recently, the work of Haimovici et al. 2013 demonstrated, using the Greenberg-Hastings model on the structural network defined by the Human Connectome, that the brain functional dynamics observed experimentally can be replicated, just by tuning the model to a region near its critical point.

In this work, we also use the Ising model to describe the macroscopic activity in the anatomical regions of the structural network defined by the Human Connectome (HC). However, different to previous studies, we propose the use of a message-passing algorithm called Belief Propagation (BP) to simulate the actual transmission of information in the brain. After the seminal introduction of BP in information theory (Pearl, 1988), this algorithm has been widely used in fields such as inference for Bayesian and Neural networks, error correcting codes, mobile communication, and probabilistic image processing, since it computes exact marginals distributions in acyclic graphs and approximate solutions for more general graphs, for which many variants have also been developed (Nishiyama & Watanabe, 2009).

Recently, a more general definition that unifies many types of message-passing algorithms on complex graph neural networks, has been named neural message passing, finding first applications in quantum chemistry and machine learning (Gilmer, Schoenholz, Riley, Vinyals, & Dahl, 2017; Lanchantin, Sekhon, & Qi, 2019). In neuroscience, there is also a recent interest for using BP as a model for neuronal processing within cortical microcircuits, related to the self-criticality of brain’s Bayesian computations (Friston, Parr, & de Vries, 2017) and for making inference in biological networks (Parr, Markovic, Kiebel, & Friston, 2019). In physics, the BP dynamics was introduced as a computationally efficient alternative to Monte-Carlo methods on Random Networks by Yedidia (Yedidia, Freeman, & Weiss, 2003). The fixed point solutions of these algorithms can be identified with the saddle point solution of the Bethe approximation of the corresponding free energy of the system (Yedidia et al., 2003). In this case, the dynamics followed by the algorithm is seen as an empirical strategy to reach the stationary states of the system, independently of the real (usually stochastic) dynamics defining the behavior of the system. Therefore, there has been only rare discussions about the correspondence between the dynamics of the algorithm and the dynamics of the physical system (Lage-Castellanos, Mulet, & Ricci-Tersenghi, 2014).

In this work, we modify this perspective by assuming that the actual mechanism of interaction between brain macroscopic regions is well described by the basic rules of the BP algorithm. The intuition behind this idea arises from the connection established a few years ago by Ott and Stoop (Ott & Stoop, 2006) between the continuous dynamics of a network of Hopfield neurons and the BP. This means that we can interpreted the BP application to a neuronal network as an estimate of the macroscopic electrical activity (voltage-firing rate) of every node in the HC by modeling their interaction with exchanging messages. In particular, in the absence of external stimulation, where initial activations can be random, the BP algorithm would predict the spontaneous functional brain state emerging from the structural properties of the network, i.e. due to the brain architecture.

Another message-passing algorithm, the Susceptibility Propagation (SP), allows the determination of long-range correlations between the state (activity) of nodes that arise around the critical point of the network (Mora, 2007). We will interpret these correlations as predictors of the functional connectivity mediated by the electrophysiological activity of the brain. Despite the lack of a direct validation for these connectivity matrices, we can explore their network properties and compare it with experimental data. In particular, we explore their modularity structure and show that the modules obtained from our correlation matrices are very similar to the resting-state Networks (RSN) that have been previously found by spatio-temporal Independent Component Analysis (ICA) of resting-state fMRI data (Power et al., 2011; Sporns & Betzel, 2016).

Finally, we show here a preliminary application of the model to studying the changes in the resting-state functional activity/connectivity -obtained with the message-passing algorithms- when the structural network presents different anomalies or disruptions. In particular, we explore Alzheimer’s Disease and lesions of the Corpus Callosum, by introducing in the HC changes consistent with studies of anatomical connectivity in subjects with these conditions.

## 2. Methods

### 2.1. Human Connectome

The combination of non-invasive imaging techniques (e.g., magnetic resonance imaging MRI and diffusion tensor imaging DTI) allows the estimation of anatomical connections throughout the human brain. Particularly, DTI provides information about the orientation of the tracts of myelinated fibers in the white matter of the brain, which allows estimating the connection paths between different areas of the brain (**Fig. 1**) (Buxton, 2009; Sporns, 2011). Although there are fundamental limitations in the diffusion images that remain unsolved, (for example, they are unable to determine the direction of the influence of connections, their inhibitory or excitatory nature and the presence of tracks bifurcations (Iturria-Medina et al. 2007), this technique is currently the most accepted option to characterize the anatomical connections.

The anatomical network of the human brain at a macroscopic spatial scale is known as the Human Connectome (HC), which was defined in parallel by Olaf Sporns in (Sporns, Tononi, & Kötter, 2005) and by Patric Hagmann in his PhD thesis (Hagmann, 2005). The nodes correspond to a set of cortical regions in both hemispheres of the brain, and the strength of the connection can be defined in several ways, for example, by counting the number of fibers that pass through the voxels that connect two regions, normalized by the size of the regions involved (Hagmann, 2005; Iturria-Medina et al., 2007). The structural network of the brain is then described by an anatomical connectivity matrix, that has $N = 998$ regions of interest with an average size of $1.5\ \mathrm{cm}^2$. These regions belong to 66 larger anatomical areas (Hagmann et al., 2008).

### 2.2. Ising model for brain regions in the HC in criticality

To simulate the critical functional state of the HC, we modeled brain activity using a disordered ferromagnetic Ising model. The binary Ising variables $s_i \in \{1, -1\}_{i=1,..,998}$ represent the functional state of each of the 998 regions of the HC (nodes), i.e. each region is modeled as being electrically "active" ($s_i = 1$) or "non-active" ($s_i = -1$). Each pair of regions ($i$ and $j$) interacts with a strength $J_{ij}$, defined by the corresponding element of the HC connectivity matrix (Hagmann et al., 2008). In our case, all $J_{ij} > 0$ since the structural network only measures a general strength of the connection (and this is why the model correspond to a ferromagnetic system). When there are not direct connections between regions $i$ and $j$, the connection is set to 0 ($J_{ij} = 0$). Then, for a specific configuration of states $\{s_i\}_{i=1,..,998}$, the interaction occurs only among the first neighbors according to the Hamiltonian:

$$\mathcal{H}\{s_i\} = -\sum_{\langle ij \rangle} J_{ij} s_i s_j - \sum_i h_i s_i \quad (1)$$

where $h_i$ represents the influence of an external field on the state of the $i$-th node, which is interpreted as the influence of external stimuli on the activation of the corresponding region. See Section J of the Supplementary Material for a Glossary and the mathematical notation used in this work. For the Ising model, the thermodynamic properties (i.e. the macroscopic characteristics) of the system are obtained from the statistical sum or partition function:

$$Z(\beta) = \sum_{\{s_i \pm 1\}} e^{-\beta \mathcal{H}(s_i)} \quad (2)$$

Where $\beta$ corresponds to the inverse of the temperature $T$ in ferromagnetic systems, where it controls the criticality of the system. For instance, in a grid of spins, higher temperatures imply that thermal fluctuations are more relevant than the interactions defined by $\mathcal{H}\{s_i\}$, spins are uncorrelated (some spins point up and some down randomly) and the net magnetization of the system is zero (paramagnetic state). At low temperature, thermal fluctuations may be neglected and the spins essentially organize trying to minimize $\mathcal{H}\{s_i\}$. When all interactions are positive (all $J_{ij} = J > 0$, i.e. pure ferromagnetic case), they will all point in the same direction, either up or down. In very general conditions, for example, system dimension larger than 1, a critical temperature $T_C$ that separates both phases can be determined. Right at $T_C$ the system is critical, in the sense that the correlation length between the spins diverges, and an infinite cluster of connected spins dominates the behavior of the system. In our case, the temperature is interpreted as a control parameter that modulates the global strength of connection between brain regions (higher $T$ implies lower $\beta$ which means weaker connections and vice versa). For $T \gg T_C$, the exchange of information is small, and the activations of anatomical regions are independent and there will be no network patterns. For $T \ll T_C$, the strong positive connections will imply that all regions are active (or inactive) and the brain cannot transport any new information on long range scales. None of these scenarios are found in real measurements of brain activity in a healthy human brain. Only near the critical point $T_C$, due to long-range correlations, the system exhibits complex networks patterns, characteristic of brain activity.

In practice, $\beta$ is analogous to a regularization parameter in an optimization problem, whose optimal value (the critical point) is not easy to estimate and usually needs the evaluation of heuristic information criteria. Fortunately for specific models, statistical physics provides a well-defined procedure to estimate the critical temperature by looking to the divergence of the susceptibility of the system (the variance of the magnetization over time) (Dorogovtsev, Goltsev, & Mendes, 2002). Since the Human Connectome is a network of finite size, the critical value of $\beta$ will be found as the one corresponding to the maximum (global peak) of the susceptibility of the system (Marinazzo et al. 2014), computed through the Metropolis algorithm (Newman & Barkema, 2001) or alternatively BP.

In this study we model resting-state activity, thus we will always consider that $h_i = 0$. The brain regions are connected only by excitatory interactions that work in both directions ($J_{ji} = J_{ij} > 0$). These assumptions are obviously a simplification of the reality. On one hand, the brain is never really "at rest" and many internal processes can have influences similar to external stimulation. On the other, it is well-known that there are many neuronal interactions based on inhibitory synapses. Unfortunately, current tractography methods to determine the connections among the cerebral regions cannot distinguish between inhibitory and excitatory ones. Likely, both types of connections can coexist within tracks of nervous fibers that contain many axons. Similarly, the current spatial resolution of the HC makes impossible to know the directionality of anatomical connections. However, using only excitatory connections that work in both directions may be a valid assumption at the large spatial scale of the anatomical network defined by the HC. Indeed it has already been shown (Hagmann et al., 2008) that macroscopic long-range connections are mainly excitatory while inhibitory synapses are more short-range.

### 2.3. Activation map using Belief Propagation

The use of an Ising model for the activity of macroscopic brain regions, have allowed the study of the potentiality of the anatomical network to create complex patterns of functional activity (Marinazzo et al., 2014), and the relationship between structure and functionality (Deco et al., 2012). Here, we intend to find the spatial functional activation map that derives from the HC by modeling the dynamics with the Belief Propagation (BP) algorithm (Yedidia et al., 2003).

Belief Propagation consists of a group of fixed-point equations for the state variables of the nodes in a network, which can be derived considering that the interaction between nodes is mediated by messages that are sent from each node to its neighbors. The message $m_{ji}$, sent from node *j* to node *i* depends on the state of both nodes $s_j$ and $s_i$, the strength of the anatomical connection between them $J_{ji}$, and the messages that were sent to node *j* from all its neighbors but *i*, in the previous iteration of the algorithm. Mathematically, it takes the form:

$$m_{ji}^{t_d+1}\left(s_i\right) = k\sum_{s_j} e^{\beta\left(J_{ji}s_j s_i + h_j s_j\right)} \prod_{l\in N(j)\setminus i} m_{lj}^{t_d}\left(s_j\right) \tag{3}$$

where $t_d$ is an index indicating iteration time; $N(j)\backslash i$ denotes the set of all neighbors of node $j$ without $i$; $\beta$ is the control parameter; $h_j$ is the strength of the influence of an external field on $j$; and $k$ is a normalization constant determined by the condition $m_{ji}^{t_d}(1) + m_{ji}^{t_d}(-1) = 1$. In this equation, recall that we will use $h_j$=0 for all nodes, the state of each node can only take two values $s_j = \pm 1$ for the Ising model and the term $J_{ji}s_j s_i$ defines the interaction between the two neighboring regions. Starting from a random distribution of states in each node, this algorithm could converge to a stationary pattern of activation, depending on the values of the interactions and the control parameter.

This definition of messages in the BP algorithm, allows to find a direct relation with the dynamical equations of the computational model of a Hopfield neural network (Ott & Stoop, 2006). This neural network consists in a large set of interconnected neuronal units, whose activity in terms of firing rate or electric potentials is approximately analogous to the behavior observed in the axons of neurons in the brain. The dynamical equations for the collective properties of the network are based on the concept of energy minimization. The equation of the continuous dynamics of the Hopfield network is summarized in (Ott & Stoop, 2006):

$$\frac{dv_i(t)}{dt} = -v_i(t) + f\left(\sum_l w_{li} v_l(t)\right) + \beta K_i(t) \tag{4}$$

where $v_i$ describes the activity of the neuron $i$ (e.g., the membrane potential); $f(x)$ is a nonlinear activation function; $w_{li}$ is the connection weight between neighboring nodes, which for the Hopfield model is symmetrical, $w_{li} = w_{il}, \forall l, i$; the parameter $\beta$ also controls the relative influence of connection weights vs. the external field (physiological noise), represented by $K_i(t)$.

The connection of this expression with the BP messages in Eq. (3) can be shown by using a nonlinear re-parameterization to new messages $n_{ji}$ that do not depend of the states of the nodes $s_i = \pm 1$ (see details in Section C of the Supplementary Material):

$$\text{Tanh}\, n_{ji} = m_{ji}(+1) - m_{ji}(-1) \tag{5}$$

Then, after convergence of messages given in Eq. (3), we compute the new messages in Eq. (5) and the brain activation (local magnetization[1]) $m_i$ of node $i$ is calculated by:

[1] Note that $m_i$ -with a simple sub-index- represents the activation of region $i$, while the $m_{ij}$ -with two sub-indices- represents the message sent from region $i$ to $j$. Notation is explained in Appendix H of the Supplementary Materials.

$$m_i = \tanh\left[\beta h_i + \sum\nolimits_{l\in N(i)} n_{li}\right] \quad (6)$$

Details of the derivation of this expression from the BP formulation are given in Section A of the Supplementary Material. The pseudo code for the BP algorithm is given in Section B of the Supplementary Material.

In the translation of Eq. (3) into Eq. (4) after the first step represented by Eq. (5), we can identify the new messages as $n_{ji} = w_{ji}v_j^i$. This allows to propose a physiological meaning to the messages of the computational model: the message ($n_{ji}$) sent by $j$ to $i$ represents the electrical activity ($v_j^i$) of the region $j$ on $i$ weighted by the synaptic connection between them ($w_{ji}$).

Finally, the activation (local magnetization) $m_i$ is identified with the macroscopic electrical activity $v_i$ of region $i$. This relation is guaranteed as long as the connection weights $w_{ij}$ and the external fields $h_i$ are relatively weak. In addition, it is also required that each region receives a large number of inputs, i.e. the degree of the network should be high. These approximations are reasonably realistic in our model because the external fields $h_i$ are equal to zero and the effect of a single synapse (connection) in the HC is typically small compared to the total number of synapses that enter a brain region (Ott & Stoop, 2006).

### 2.4. Determining the long-range correlations using Susceptibility Propagation

Experimental studies on brain functional connectivity has shown that distant elementary structures are functionally correlated (Michael D Greicius et al., 2003). It means that elements without a direct structural connection (fiber tracts), present a temporal synchronicity in the dynamics of their hemodynamical BOLD signals. Unfortunately, BP does not provide a direct path to estimate the correlation between non-interacting spins. In recent years, another message-passing algorithm, named Susceptibility Propagation (SP), has been built on top of BP and allows the computation of these zero-lag correlations in a simple and efficient way (Aurell, Ollion, & Roudi, 2010; Charles Ollion, 2010; Marinari & Van Kerrebroeck, 2010; Mora, 2007). Since the susceptibility[2] of each pair of nodes is defined as the variance of its states, the SP determines the covariances $\chi_{ij}$ between nodes' activations given the known interaction strengths $J_{ij}$ and the external local fields $h_i$ in each region. For this algorithm, two messages sent from region $i$ to $j$, given the influence of the local field of region $k$, ($g_{i\to j,k}$ and

[2] This local susceptibility refers to the variance of the state of the spin in time, which refers to the long-range correlation between elements of the network. Do not confuse with the system susceptibility defined above to calculate the critical point.

$v_{i\to j,k}$), are defined as the derivatives of the BP messages, conveniently rewritten using the log-likelihood notation ($h_{i\to j}$ and $u_{i\to j}$)[3] (Charles Ollion, 2010):

$$h_{i\to j} = \beta h_i + \sum_{l\in N(i)\setminus j} u_{l\to i} \qquad u_{i\to j} = \tanh^{-1}\left[\tanh\left(\beta J_{ji}\right)\tanh\left(h_{i\to j}\right)\right] \tag{7}$$

$$g_{i\to j,k} = \beta\delta_{ik} + \sum_{l\in N(i)\setminus j} v_{l\to i,k} \qquad v_{i\to j,k} = g_{i\to j,k}\tanh\left(\beta J_{ij}\right)\frac{1-\tanh^2\left(h_{i\to j}\right)}{1-\tanh^2\left(u_{i\to j}\right)} \tag{8}$$

where $\delta_{ij}$ represents the Kronecker delta ($\delta_{ij} = 1$, for $i = j$; $\delta_{ij} = 0$, otherwise). The messages given by Eq. (7) and Eq. (8) are initialized randomly. Once the messages corresponding to BP Eq. (7) converge, the activations (local magnetizations) are calculated according to:

$$m_i = \tanh\left[\beta h_i + \sum_{l\in N(i)} u_{l\to i}\right] \tag{9}$$

Which is equivalent to Eq. (6) but expressed in terms of the messages $u_{i\to j}$. Then the new messages Eq. (8) are updated only at the critical point. Finally, the long-range correlations are determined from:

$$\chi_{ij} = \left(\beta\delta_{ij} + \sum_{l\in N(i)} v_{l\to i,j}\right)\left(1-m_i^2\right) \tag{10}$$

Details of the derivation of this expression can be found in the Section D of the Supplementary Material. The pseudo code for the SP algorithm is given in Section E of the Supplementary Material. It is important to note that -although it is not straightforward to see from equations- the susceptibility in Eq. (10) leads to a symmetrical matrix. In statistical mechanics, this is called the correlation function and it is bounded from -1 to 1 since all probability distributions (messages) are normalized and states are defined as ±1. However, this does not mean that it is normalized in the sense of the classical statistical correlation, i.e. divided by variances. Although using the typical normalization by variances has been proposed in the context of the Bethe's approximation of an Ising model as a trick for improving inference (Ricci-Tersenghi, 2012), we here follow the more traditional procedure of ignoring variances (diagonal values of the matrix formed by elements $\chi_{ij}$ are set to 0) after convergence of SP.

[3] Similar to the previous section, here $h_i$ -with a simple sub-index- represents the external field in region $i$, while $h_{i\to j}$ represents the message sent from region $i$ to $j$, rewritten in log-likelihood notation.

## 2.5. Validation of the model

An important confound that must be ruled out in any study of models defined in a network, is whether the results obtained are specific for the network of interest, can also arise from any random network with similar properties. This can be addressed by defining proper null models and compare results obtained in these networks with those of the original network (Rubinov & Sporns, 2010; Sporns, 2018). In this work we constructed two random connectomes that represent the null hypotheses for our model on the Human Connectome (HC), whose connectivity matrix is shown in **Fig. 1a**. They are defined as follows:

Random Connectome 1 (RC1): This network preserves the same number of nodes and connections of the HC, but nodes are randomly reconnected. This null model allows testing whether the results obtained using the HC are due only to the local characteristics of the network, such as the degree of each node, independently of its topology. The corresponding connectivity matrix is shown in **Fig. 1b**.

Random Connectome 2 (RC2): This network preserves the topology of the HC, but the weights of the connections between each pair of nodes are randomly reassigned. This null model allows testing whether the results obtained using the HC are due only to the topology and not to the particular distribution of connection values. The corresponding connectivity matrix is shown in **Fig. 1c**.

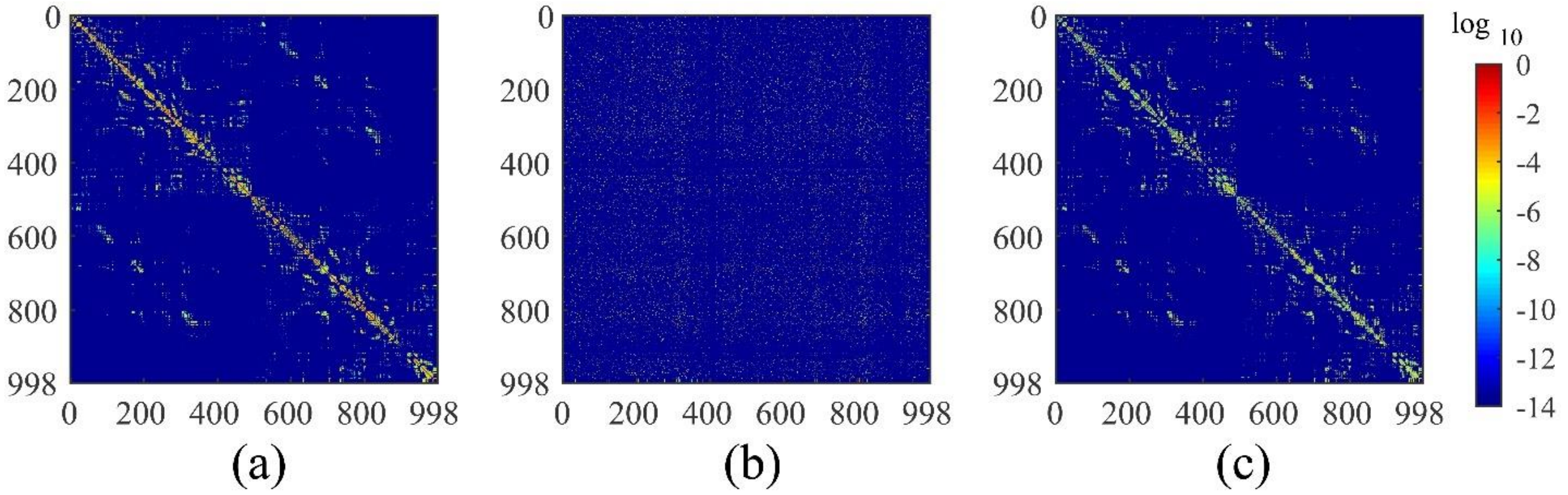

**Fig. 1.** Anatomical connectivity matrices where connection weights are represented in a logarithmic color code, for a better visualization. a) Anatomical connectivity matrix of the Human Connectome. b) Anatomical connectivity matrix of the random connectome RC1, which represents a null network with random topology but keeping the basic local properties of the HC. c) Anatomical connectivity matrix of the random connectome RC2, which represents a null network with the same topology of the HC, but with the connection weights randomly reassigned to existing connections.

Both null networks were generated with the Brain Connectivity Toolbox (Rubinov & Sporns, 2010), which provides multiple functions of network manipulation.

Since the magnetization in each node is interpreted as the neuronal activity (activation) according to the Hopfield model, one must remember that in principle this corresponds to the macroscopic electrical activity of large neuronal populations (as defined by specific nodes in the HC). Currently experimental techniques do not allow the direct measurement of such a magnitude; therefore, we here follow a practical approach and compare the maps provided by the model with those obtained experimentally from measurements of the blood oxygenation level dependent (BOLD) signal, which have been the commonest data for studying functional resting-state brain networks.

In this sense, it is important to note that the BOLD signal is only an indirect measurement of the electrical activity of the brain. From the signal processing point of view, the first can be considered a temporal convolution of the second (Daunizeau, Laufs, & Friston, 2009; Luessi, Babacan, Molina, Booth, & Katsaggelos, 2011; Rosa, Daunizeau, & Friston, 2010). In transient phenomena this convolution cannot be ignored. However, in this paper, we are dealing with a stationary functional resting state, where it is reasonable to ignore this temporal relationship and consider the BOLD signal as a temporal average of the electrical activity (proportional to the variance of the neuronal electrical activity) (Trujillo-Barreto NJ, Martínez-Montes E, Melie-García L, & Valdés-Sosa PA, 2001; Valdes-Sosa et al., 2009). In other words, as a representation of lower temporal resolution of the brain's electrical activity.

In addition to the visual qualitative comparison from images, we performed a Receiver Operating Characteristic (ROC) analysis to compare the results obtained with the implementation of the BP on the three networks (HC, RC1 and RC2) with a thresholded experimental map of the Default Mode Network (DMN) of the brain. We report the area under each ROC curve (AUC) as a general measure of the similarity between pair of maps (Fawcett, 2006; Grova et al., 2006; Swets, 1988). A bootstrapping method was used to generate surrogate ROC curves that allowed the assessment of variability of the AUC measure, as well as the computation of confidence intervals, for a proper statistical comparison between the models (Fawcett, 2006).

To quantitatively evaluate the relevance of the correlation matrices obtained with the SP algorithm, we explored the clusters or modules derived from them, which would suggest a kind of functional organization based on the anatomy of the brain. The clustering was done using the *modularity_und* function of the Brain Connectivity Toolbox (Rubinov & Sporns, 2010), by tuning the gamma parameter to always obtain 12 modules (after rejecting any cluster having less than 38 nodes), to approximately match the number of clusters found in the empirical RSNs. An additional cluster is formed with the unassigned nodes, but it is not taken into account for subsequent qualitative and quantitative comparisons as it does not reflect organization of the connectivity matrices.

The community structure obtained from these clusterings were quantitatively compared using the Normalized Mutual Information (NMI), which is a non-linear measure of the similarity between two partitions, based on information theory (Danon, Díaz-Guilera, Duch, & Arenas, 2005). See Section F of the Supplementary Material for a more detailed presentation of this measure. We then analyzed the correspondence of the communities or modules derived from estimated long-range connectivity matrices with the clusters of functional connectivity appearing in the experimental resting-state brain networks (RSN).

The experimental activations of the RSNs were represented with the masks described in (Beckmann, DeLuca, Devlin, & Smith, 2005), which were translated to the Human Connectome of 998 regions in (Haimovici et al., 2013). For this, we first computed -for each partial RSN separately- the percentage of overlapping nodes between each module and the RSN and ranked modules according to the highest overlapping. The module with maximum overlapping (i.e. the one that ranked first) was assigned to the RSN. In those cases where the same module ranked first in two RSNs, we left it assigned to the RNS with highest overlapping and for the other RSN we assigned the module that ranked second. This matching procedure did not take into account the unassigned nodes. It was done for all connectivity matrices obtained with the SP and for the anatomical HC matrix in order to facilitate visual and quantitative comparisons explained above. With this matching, it is easy to create confusion matrices for evaluating the similarity of each clustering to the experimental RSNs (used as gold standard), with typical measures such as Accuracy, Sensitivity (recall), Precision and F1 score (Fawcett, 2006). Finally, the representation in the brain of assigned modules were visually compared with the maps of the experimental RSNs.

### 2.5.1. Preliminary exploration of the usefulness of the model

Within the framework of the proposed model, we can then predict the functional map corresponding to the DMN in different pathologies or diseases, by introducing changes in the structural HC (e.g. disrupting specific connections or varying the connectivity strength) as reported in the literature (Griffa, Baumann, Thiran, & Hagmann, 2013). We chose two examples that are widely studied and imply changes in the HC that are relatively easy to reproduce, such as Alzheimer's Disease (AD) and the absence of the Corpus Callosum (e.g. by agenesis or by surgical severing).

To simulate AD in the connectome, we reduced the connection weights in the whole system by 20% ($J_{ij} = J_{ij} \times 0.8$). For the simulation of the absence of the Corpus Callosum (CC), all weights for connections between the right and left hemisphere were set to zero ($J_{ij} = J_{ji} = 0$, if $i$ and $j$ belong to different brain hemispheres). In this work, we explored the behavior of functional maps and connectivity modules when using the critical value of the original HC, in order to simulate the immediate functional alterations that may appear from disruptions of the anatomical network. It is easy to see that computing the critical value for the altered AD network will only re-scale all connections and both the activity and the connectivity will give exactly the same solution as the original HC. The case of CC could be more interesting and they will both be discussed.

## 3. Results

### 3.1. Functional maps emerging from the structural brain network in criticality

For each of the structural networks studied (the HC, and the two null-hypotheses networks RC1 and RC2), the critical control parameter was determined as the value that led to the maximum susceptibility of the system (as evaluated in a grid of 100 values, see Section G of the Supplementary Material). Using these critical parameters, we ran BP and SP for each anatomical network starting from 100 randomly chosen initial conditions and without external stimulus ($h_i = 0, \forall i$) to simulate the resting state. Convergence was found to the same final solution (activation map or connectivity matrix) in more than 95% of the repetitions for every case, indicating that each system has a global attractor. From BP we obtain a vector of 998 components whose values range between 0 and 1, reflecting the spontaneous activation of the 998 regions of the connectome. The vector is then represented as a color-coded map overlaid to a template T1-weighted MRI image in the standardized Montreal Neurological Institute (MNI) space, using the software Neuronic Tomografic Viewer. All estimated maps were thresholded at 0.7 (i.e. activations below that value were set to zero) to improve the visualization. **Fig. 2a-c** show the maps obtained for the three anatomical networks. **Fig. 2d** show the map of the Default Mode Network (DMN) obtained experimentally from the typical resting-state analysis of BOLD signals (Beckmann et al., 2005), in the same template and coordinate system used to illustrate our results.

Note that only the spatial map obtained when using the HC (**Fig. 2a**) resembles the experimental DMN map (**Fig. 2d**) reported in the literature. In the case of the RC1 and RC2, activations are less symmetric and appear in non-relevant areas of the brain. The frontal activation seen in the experimental map was not reflected in any of the estimated maps, which might support the idea that this activation is related to spontaneous cognitive processing instead of just arising from the anatomical architecture.

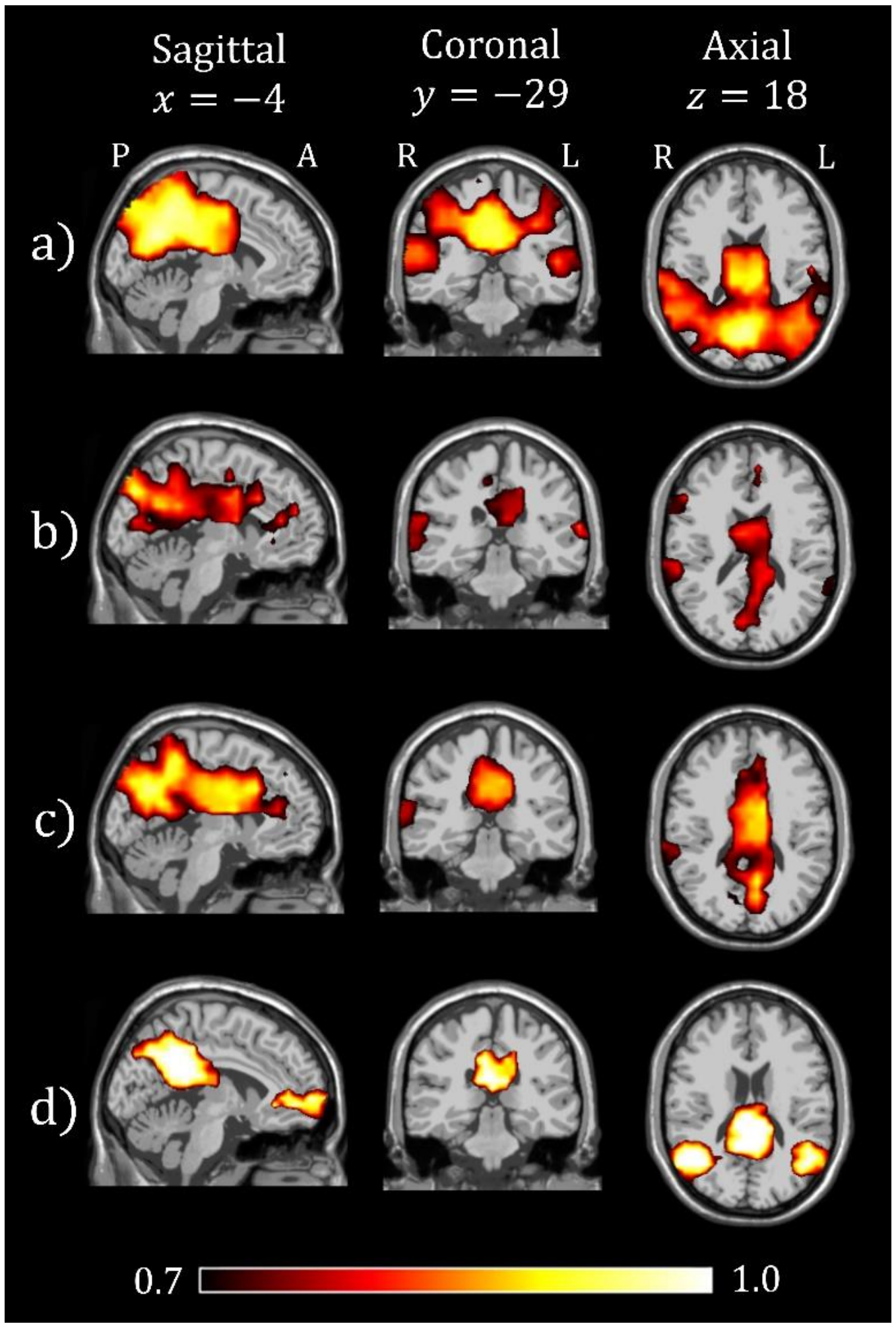

**Fig. 2.** The images (a-c) show the activation maps obtained with the BP algorithm in the Human Connectome, the Random Connectome 1 and the Random Connectome 2, respectively, interpolated in a T1-weighted MRI in the MNI standardized space. In addition, image (d) illustrates the experimental map corresponding to the brain's default mode network that was visualized in the same space using the masks described in (Beckmann et al., 2005). P, A, L, and R stand for Posterior, Anterior, Left and Right.

To perform a more quantitative comparison we plot ROC curves derived from the distributions of activity in the three maps, using the thresholded experimental DMN map as the gold standard (**Fig. 3**). We used the Matlab function *perfcurve* which, in addition to get the AUC for each curve, allows us to compute the confidence bounds for 1000 bootstrap replicas, from where a sample unpaired t-test can be performed. This measure provides a robust test of whether the comparisons are different from the experimental data. **Fig. 3** shows that the AUC of the distribution of activity in the Human Connectome is significantly higher than 0.5 ($z = 10.7$; $p < 1e - 5$) and significantly higher than the AUC for RC1 ($t = 2.02$; $df = 1998$; $p = 0.02$) although not for RC2 ($t = 0.09$; $df = 1998$; $p = 0.47$). In turn, the AUC for RC2 was significantly higher than that for RC1 ($t = 2.08$; $df = 1998$; $p = 0.02$).

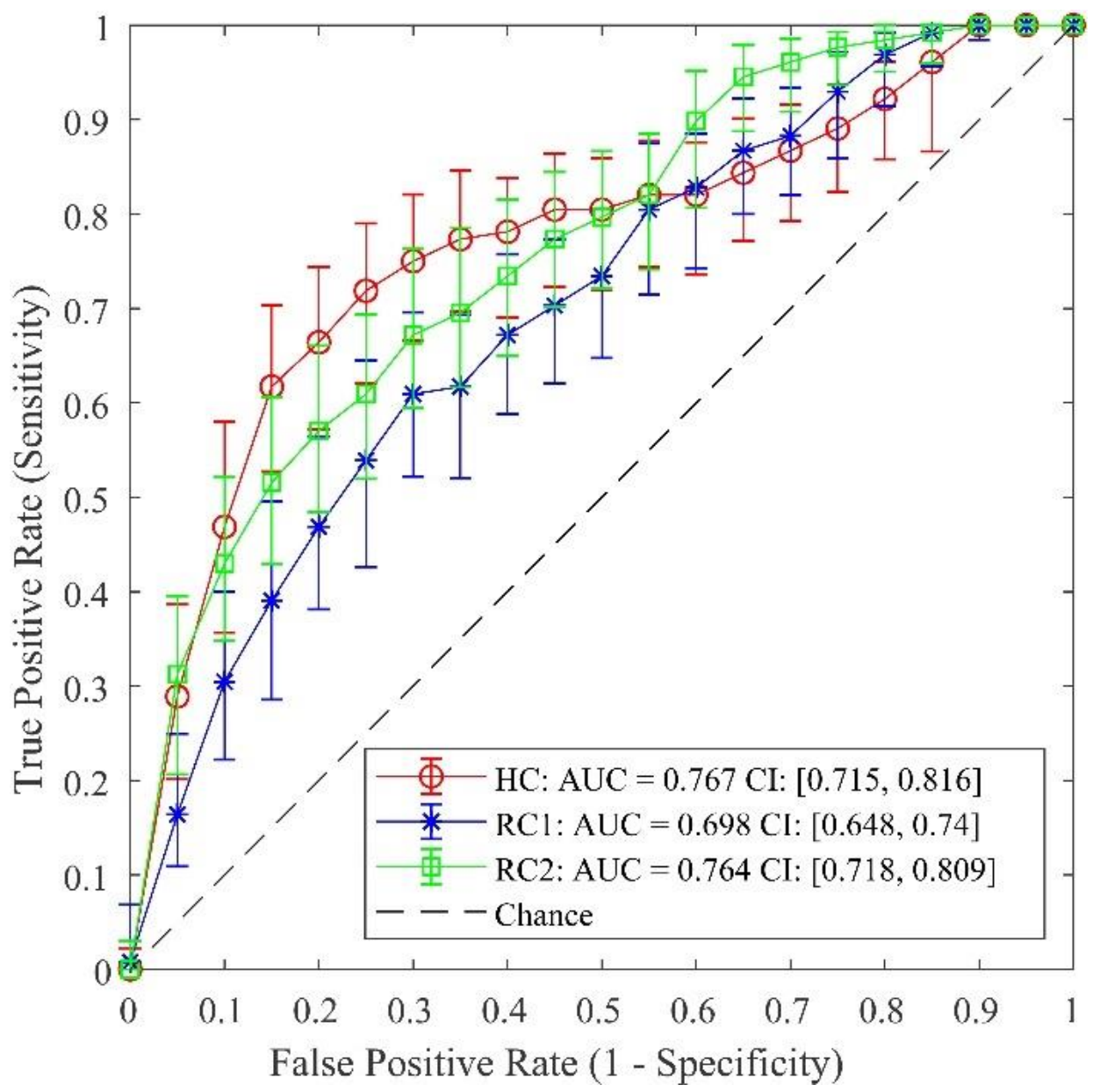

**Fig. 3.** Receiver Operating Characteristic (ROC) curves for the activation maps obtained using the BP algorithm from the HC (red circles) and the null hypotheses networks RC1 (blue stars) and RC2 (green squares). The experimental DMN was used as the gold standard. The "area under the curve" (AUC) for all cases are shown in the legend, together with confidence intervals obtained from bootstrap replicas. The plot represents the mean values across bootstrap replicas and error bars for each method.

### 3.2. Long-range functional connectivity emerging from the structural organization

The results of the SP implementation on the three networks studied are shown in **Fig. 4**. These long-range correlation matrices can be interpreted as representing the covariance of macroscopic electrical activity of each pair of brain regions. While RC1 represents a very homogeneous pattern of correlations, the correlation matrices of HC and RC2 show non trivial structures. However, the HC leads to a sparser matrix with a clearer modular structure.

**Fig. 5a-c** shows the division into modules (also known as clusters or communities) made from the three correlation matrices given in **Fig. 4**. **Fig. 5d** reflects the modules determined directly from the structural HC network, i.e. taking the original $J_{ij}$ (and not the results of SP) as a measure of the correlations between the nodes. As expected, the correlations obtained from the randomly connected network RC1 showed communities with little structure, formed by nodes with scattered positions in the brain. On the other hand, the correlation matrices obtained from HC and RC2 reflected well-defined modules formed by spatial neighbor regions, similar to the community structure directly obtained from the anatomical connectivity matrix of the HC (**Fig. 5d**). However, it can be seen that the distribution and sizes of the modules are different for the three cases, with a much higher symmetry for the clustering of the correlation matrix obtained with SP from the HC network.

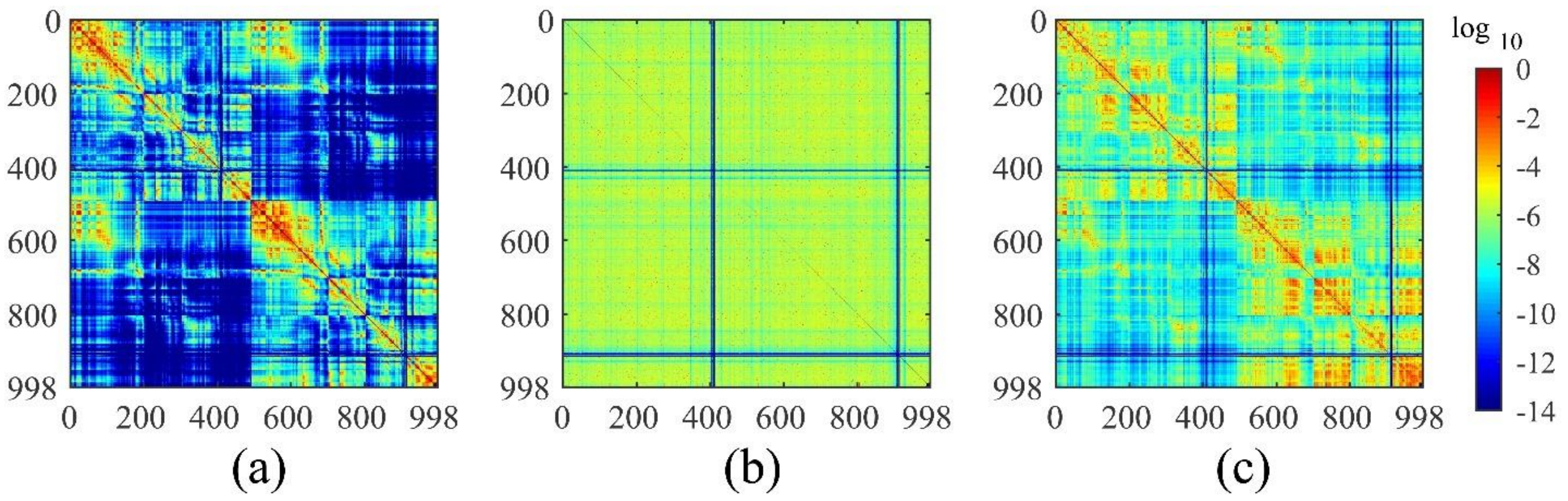

**Fig. 4.** Matrices of long-range correlations among the 998 brain regions, obtained from the implementation of the SP on the structural networks corresponding to **a)** the Human Connectome (**Fig. 1a**), **b)** the Random Connectome RC1 (**Fig. 1b**), and **c)** the Random Connectome RC2 (**Fig. 1c**).

**Fig. 6a** shows the quantitative differences between these clusterings as measured by the normalized Mutual Information (NMI), which confirmed our visual perception of the inequalities between the divisions in modules on the four networks. The smallest NMI values corresponded to the comparisons of all the networks with the network of random topology SP-RC1. The comparison among modular divisions of the functional connectivity matrices obtained with the SP algorithm from the HC and the RC2 (SP-HC and SP-RC2), and from the HC directly, offered similar values of NMI for each pair of them (0.56 – 0.57). Given that they showed similar division in spatial clusters, this result suggests that the modules obtained from the different matrices have intermediate levels of overlapping.

To explore the functional relevance of the modules extracted from these connectivity matrices, we matched modules in every clustering to each of the ten subsystems (activation patches) of experimental functional resting-states networks (RSNs). These subsystems were defined by the masks described in (Beckmann et al., 2005) and translated to the Human Connectome of 998 regions in (Haimovici et al., 2013). The RNSs subsystems are: Visual Left, Visual Right, Auditory Left, Auditory Right, Sensory-Motor Left, Sensory-Motor Right, Default Mode Network, Executive Control, Dorsal Visual Stream Left, Dorsal Visual Stream Right. **Fig. 7a** shows the behavior of each connectivity matrix (the three functional SP-HC, SP-RC1, SP-RC2, and the anatomical HC) according to Accuracy, Sensitivity, Precision and F1 score. These curves show that the modules obtained with SP-HC are in higher correspondence with experimental RSNs than those obtained from SP-RC2 and from the HC anatomical connectivity. Interestingly, the anatomical HC and the correlation matrix from null network RC2 gave very similar values of all evaluation metrics. As expected, the lack of structure in the correlation matrix obtained from RC1 led to very low values in all evaluation metrics.

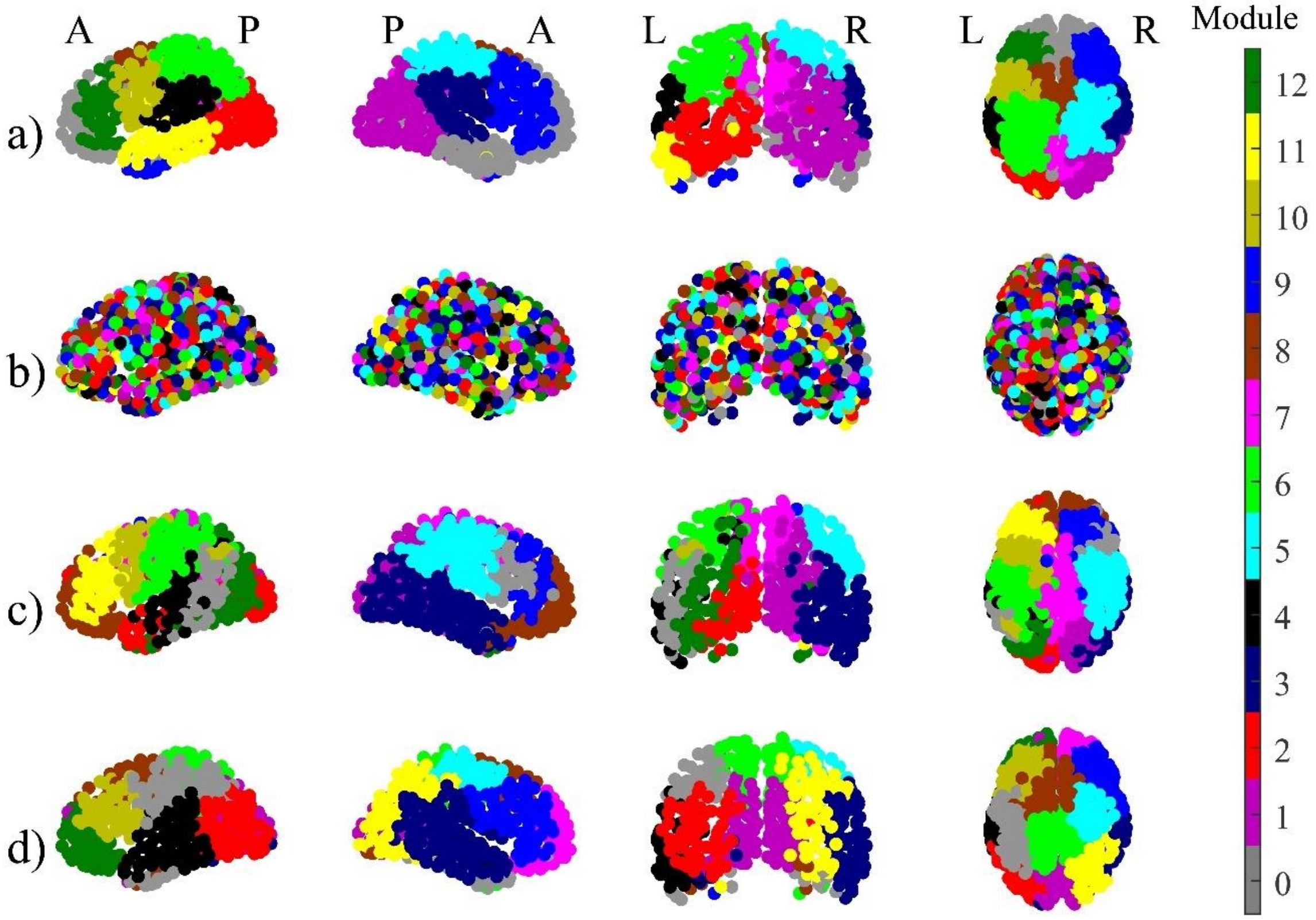

**Fig. 5.** Modularity of connectivity matrices shown in four different views (from left to right: Left, Right, Back, Top). Nodes correspond to the 998 brain regions in the HC, and those belonging to the same module are represented with the same color. Rows **a)**, **b)**, and **c)** show the clustering on the correlation matrices obtained with SP on the HC, RC1 and RC2, respectively (**Fig. 4**). Row **d)** shows the clustering on the anatomical matrix of the HC (**Fig. 1a**). The *modularity_und* function from the BCT was used, tuning the resolution parameter gamma such that the divisions in the three matrices will result in 12 modules (module 0 was then formed with all unassigned nodes). P, A, L, and R stand for Posterior, Anterior, Left and Right.

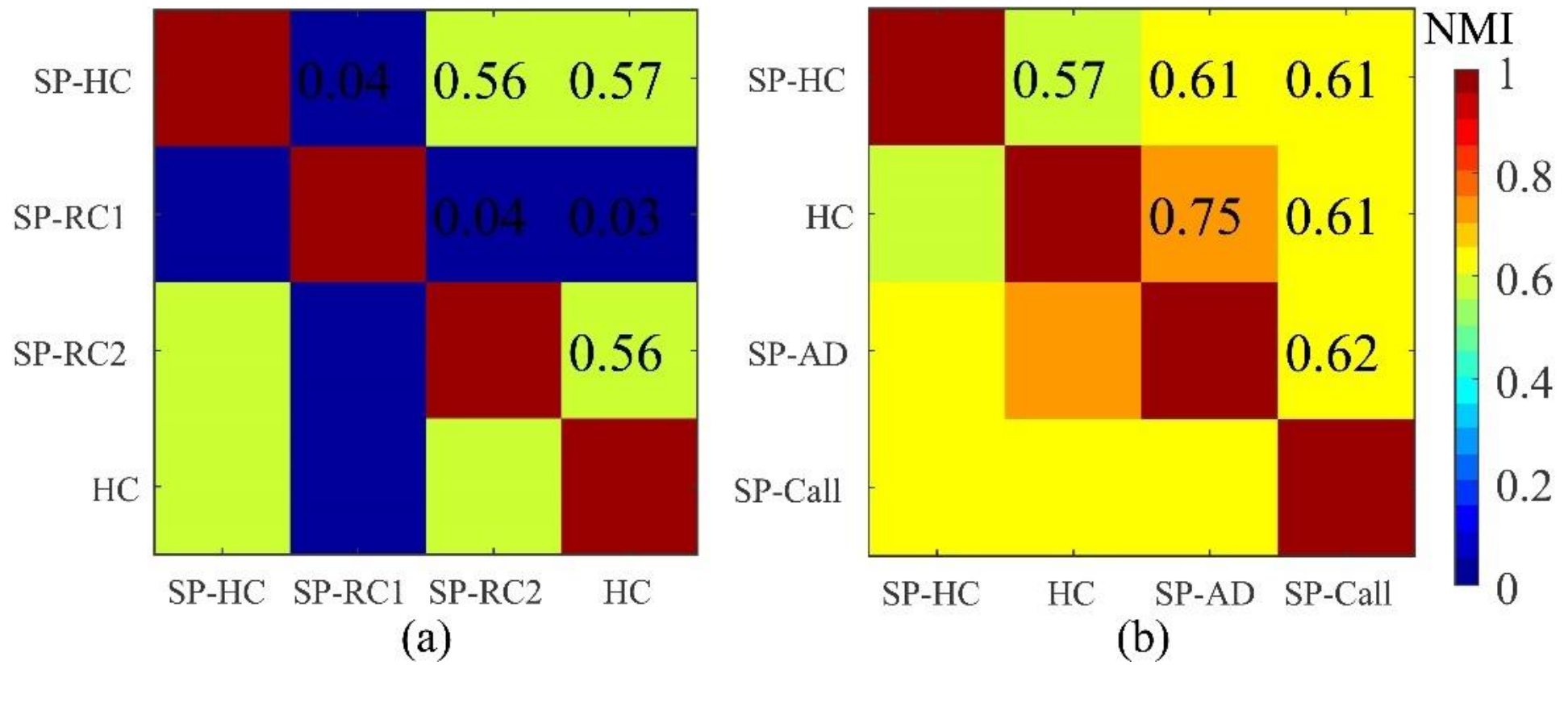

**Fig. 6.** Color-coded symmetrical matrices of the Normalized Mutual Information (NMI) for pairwise comparisons of the clustering in modules obtained from connectivity matrices. a) Comparisons among communities derived from the correlation matrices obtained with SP from: the HC (SP-HC), the null networks RC1 (SP-RC1) and RC2 (SP-RC2); as well as from the anatomical connectivity matrix (HC). b) Comparisons among the communities derived from the correlation matrices obtained with SP from the altered anatomical networks to simulate Alzheimer's Disease (SP-AD) and lesions of the Corpus Callosum (SP-CC); as well as with the same SP-HC and HC as in (a). Quantitative values of the NMI are shown over the corresponding colored pixel for an easier comparison.

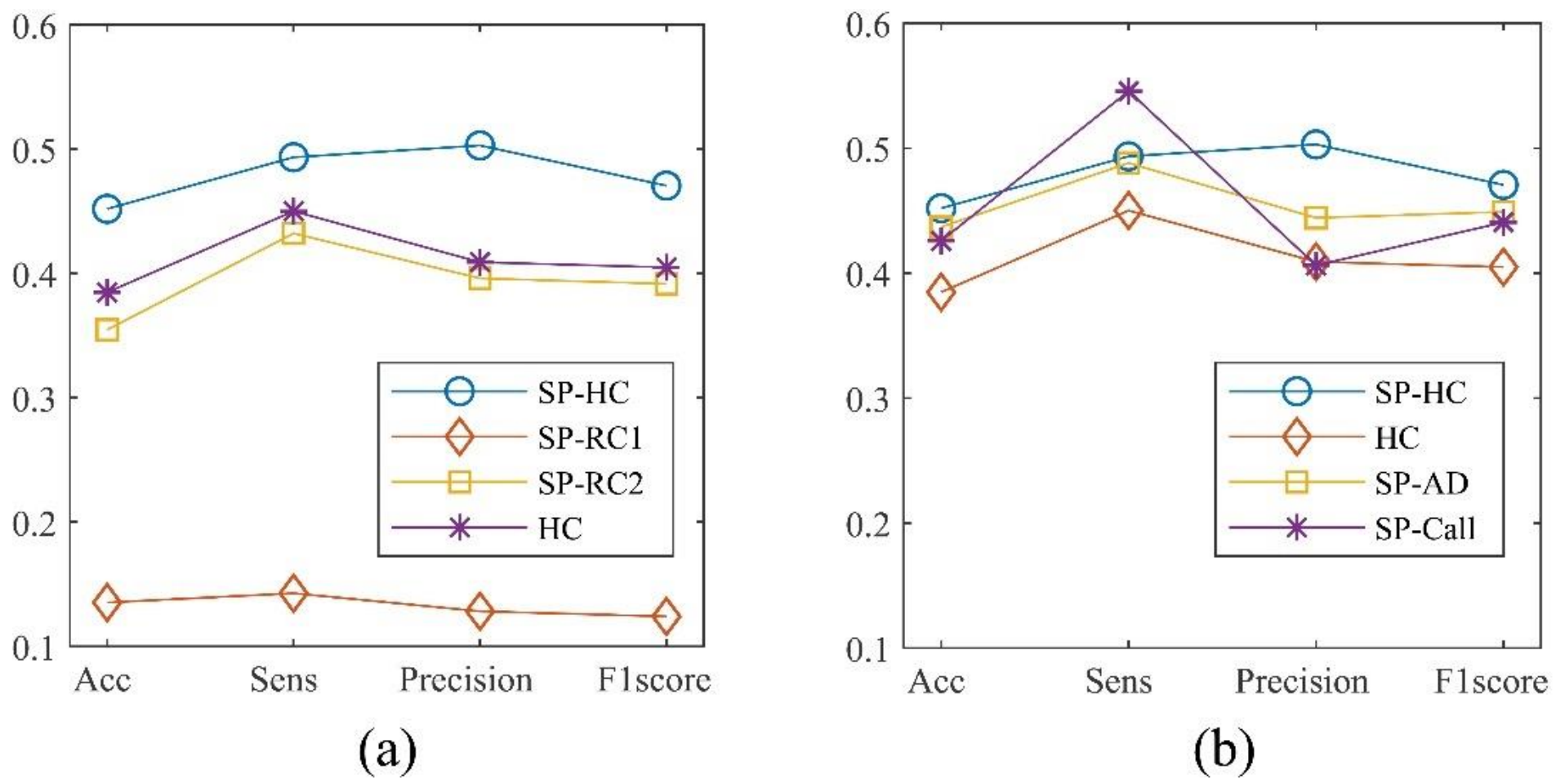

**Fig. 7.** Evaluation metrics of the correspondence between functional clusters (modules) of the connectivity matrices and a gold-standard clustering formed by subsystems of experimental RSNs. The metrics are the total (across all RSNs) Accuracy (Acc), Sensitivity or recall (Sens), Precision and F1 score. a) Values obtained for the correlation matrices obtained with SP from: the HC (SP-HC), the null networks RC1 (SP-RC1) and RC2 (SP-RC2); as well as from the anatomical connectivity matrix (HC). b) Values obtained for the correlation matrices obtained with SP from the altered anatomical networks to simulate Alzheimer's Disease (SP-AD) and lesions of the Corpus Callosum (SP-CC); as well as with the same SP-HC and HC as in (a).

**Fig. 8** represents -as brain activation maps- the modules of the connectivity matrix obtained with the SP on the HC that were matched to the ten experimental RSNs. In short, modules 1 and 2 maximally overlapped to the occipital subnetwork, the visual RSN; modules 3 and 4 resemble the auditory RSN; and modules 5 and 6 are slightly extended versions of the sensory-motor RSN. Although less clear, partial similarities were also found on more complex RSNs such as the back cluster of the DMN, (resembled by module 7), the medial activation of the executive control RSN, (overlapped with module 8), and the frontal activations of the dorsal visual stream RSN (similar to module 9 on the right and modules 10 and 12 on the left). Section H of the Supplemental Material shows similar figures with the matching of RSNs to the modules obtained directly from the HC anatomical matrix (**Fig. H2**) and from the functional connectivity matrix SP-RC2 (**Fig. H1**). They allow to visually confirm that the SP-HC is the one with the best overall match.

### 3.3. Resting-state maps and connectivity in altered anatomical networks

**Fig. 9** shows the fixed-point maps obtained with the BP algorithm on the two altered connectomes that simulates Alzheimer's Disease (AD) by decreasing all connection weights in a 20% and lesions of the Corpus Callosum (CC), by setting to 0 all connection weights between nodes of different hemispheres. In both cases, the BP was run using the same value of the critical parameter that was obtained in original conditions. In the case of AD, we found a general decrease in activations, mostly in parietal and temporal regions, which is in accordance to previous experimental fMRI studies on the characteristics of the Default Mode Network (DMN) in AD patients (Buckner et al., 2008).

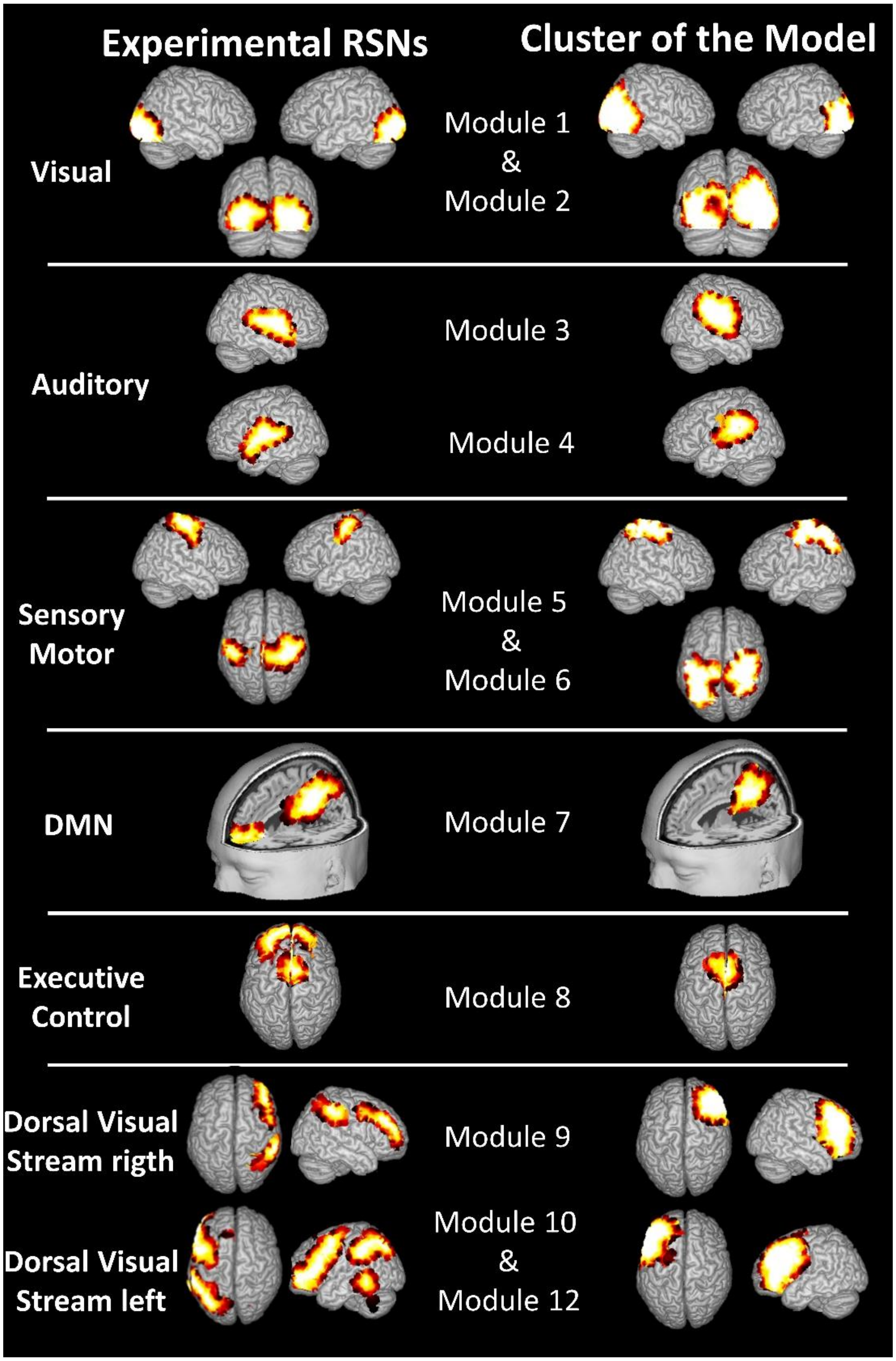

**Fig. 8.** Similarity between the modules derived from the correlation matrix (right column) obtained with SP on the Human Connectome (SP-HC, **Fig. 5**a) and the spatial maps of experimental functional resting-state networks (left column).

In the case of CC, a partial loss of interhemispheric symmetry in activations appeared, as well as smaller activations around the corpus callosum, which has also been seen in previous studies on the DMN in patients after Callosotomy (i.e. surgical severing of the Corpus Callosum) (Roland et al., 2017). Interestingly, if these maps are re-computed using the critical parameter of the disrupted connectomes, they show a high resemblance to the same maps as in normal conditions (see **Fig. I2** and **I3** of the Supplementary Material). This was expected in the case of the AD but not for the case of CC, since the topology of the network is not changed. However, there are clinical studies showing that the DMN of persons with agenesis of the Corpus Callosum or after long time of Callosotomy is very similar to DMN of healthy subjects (Tyszka, Kennedy, Andolphs, & Paul, 2012).

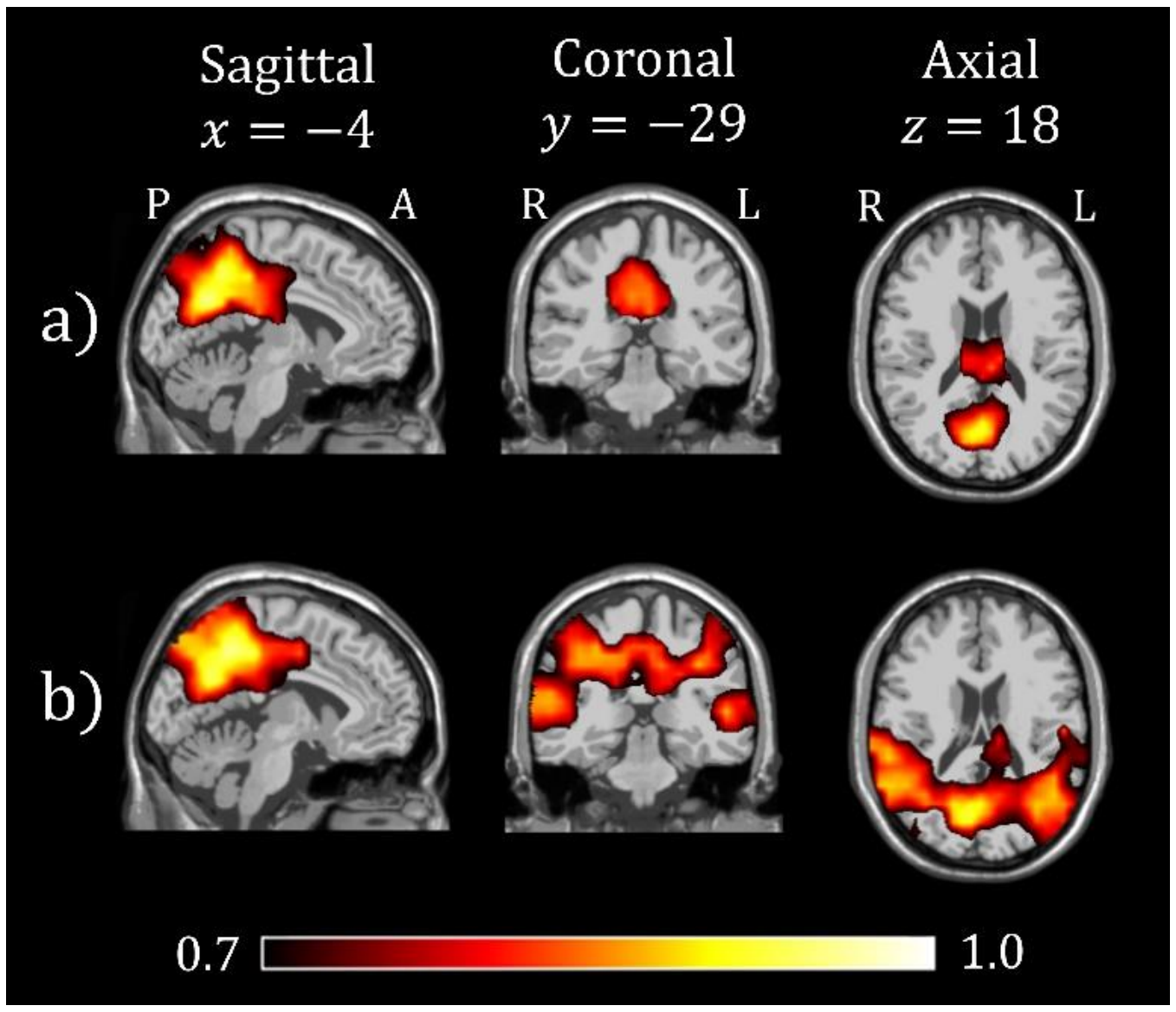

**Fig. 9.** Sagittal, coronal and axial views of the activation maps obtained with the BP algorithm from structural connectomes that simulates a) the effect of Alzheimer's Disease, where all anatomical connections' strengths were decreased by 20%; and b) lesions of the Corpus Callosum (agenesis or Callosotomy), where all connections between both hemispheres were removed. In both cases, the maps were obtained using the same critical value for the control parameter that was computed for the unaltered HC, and thresholded by 0.7. (P, A, L, and R stand for Posterior, Anterior, Left and Right).

Long-term correlation matrices of the functional activity in the resting state appearing in these altered connectomes were computed using SP, with the same critical parameter as found in the original HC. In the case of AD, the connectivity matrix showed to be sparse but still structured in clusters of connections (**Fig. 10a**). In the case of CC, the connectivity matrix showed that it is not possible to find correlations between nodes that are not connected by any path in the network (in this case, belonging to different hemispheres), but inside each hemisphere functional connections also showed a clustered structure (**Fig. 10b**). The modules obtained from clustering these matrices are shown in **Fig. 10c** and **Fig. 10d**, respectively**.** Although in both cases a clear modular division emerged, the quantitative comparison using the NMI showed that these clusterings were similarly different from the SP-HC (NMI = 0.61) and between them (NMI = 0.62), as shown in **Fig. 10b**. Interestingly, the clustering of the AD was most similar to that obtained from the anatomical HC matrix (NMI=0.75). As expected, the modules derived from the connectome simulating lesions of the CC were separated in the two hemispheres, but showed a high symmetry and larger clusters (**Fig. 10d**). This might explain an unexpected result found when studying the correspondence of modules from these two clusterings with the set of 10 experimental RSN subsystems. **Fig. 7b** shows the values of the evaluation metrics for the classification comparison using the 10 clusters of experimental RSNs as gold-standard. Although the figure suggests that the alterations of the HC leads to a poorer resemblance of the experimental RSNs, these clusterings still outperform the results from both null-model random networks and from the anatomical HC. Particularly, SP-AD have a similar behavior than the HC (which is expected, given the similarity between clusterings) but with slightly higher values. The larger modules derived from SP-CC might explain that its sensitivity is higher than that of the clustering of the SP-HC matrix, although the impossibility of resembling medial RSNs such as the DMN and the Executive Control RSN probably implied the lower Precision and total Accuracy.

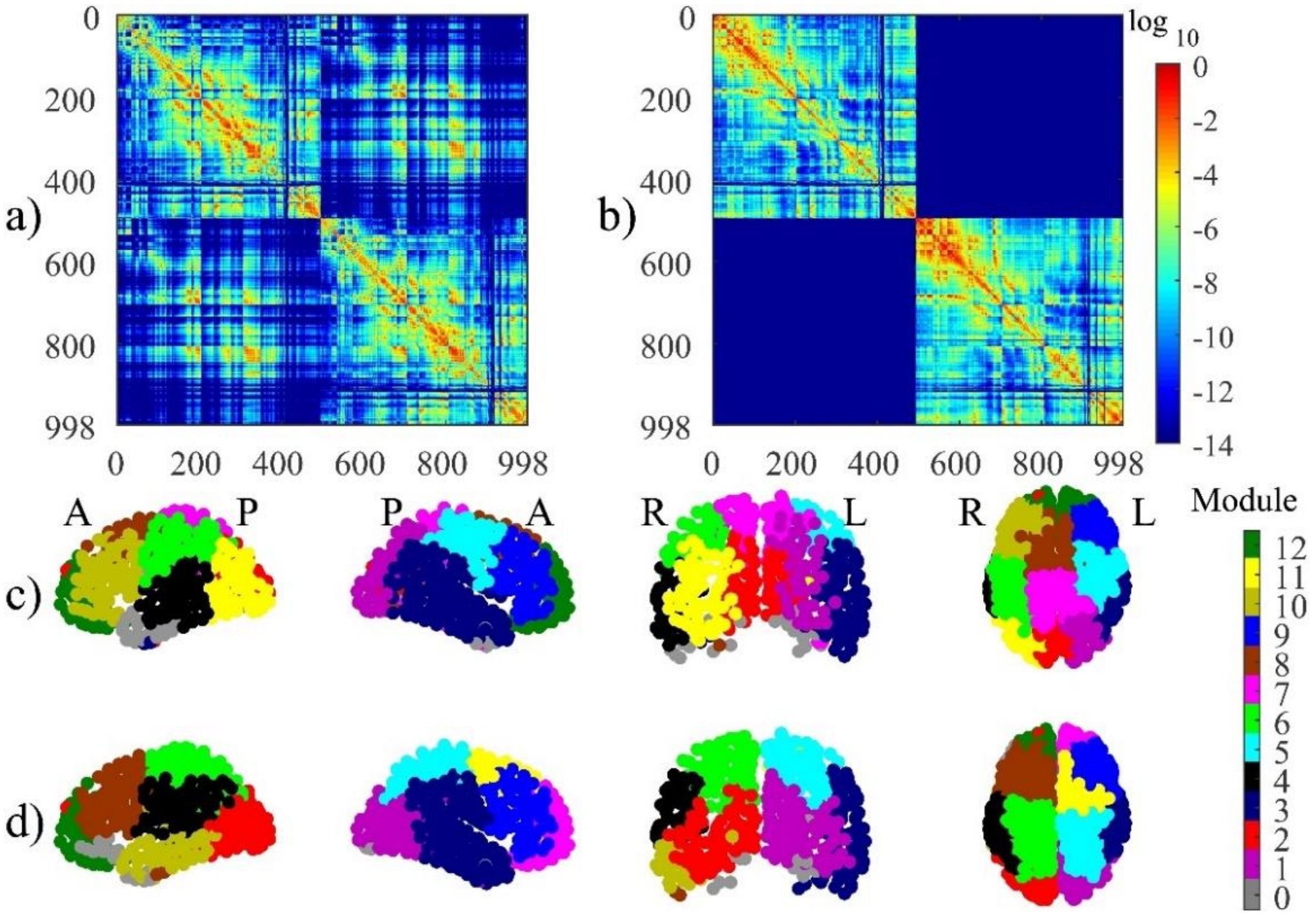

**Fig. 10.** Matrices of long-range correlations among the 998 brain regions, obtained from the implementation of the SP on the altered structural networks corresponding to **a)** Alzheimer's Disease and **b)** lesions of the Corpus Callosum. Panels **c)** and **d)** show the clustering on the matrices shown in **a)** and **b)**, respectively; in four different views (from left to right: Left, Right, Back, Top). Nodes correspond to the 998 brain regions in the HC, and those belonging to the same module are represented with the same color. The clusterings were obtained by the *modularity_und* function from the BCT, tuning the resolution parameter gamma such that the divisions in the three matrices will result in 12 modules (module 0 was then formed with all unassigned nodes). P, A, L, and R stand for Posterior, Anterior, Left and Right.

## 4. Discussion

### 4.1. Neural message passing for predicting functional activation and connectivity

In this work, we propose a neural message-passing model as a paradigm to unveil, from the brain structural connectivity, the brain functional activity at the macroscopic level. In particular we assume that brain functional activity can be modeled by allowing the nodes defined by the Human Connectome to have binary states that evolve according to a neural message-passing algorithm. This approach offers a more direct interpretation of the macroscopic electrical activity of large neuronal masses, as arising from their interaction through messages that closely follows the known biophysical rules of neuronal synapsis. In a nutshell, neurons fire bursts of action potentials whose properties (firing rate) are defined from nonlinear processing of the inputs received from other neurons. At the macroscopic level, regions will be active or non-active depending on the integration of all messages (inputs) received. The average level of activity in each region (equivalent to macroscopic potentials) is found by the difference between the probabilities of being active and non-active. To estimate the statistical correlation of the states of every pair of regions we then use the Susceptibility Propagation algorithm, which lead to a long-range correlation matrix that can indeed be interpreted as a measure of functional connectivity at the level of cortical brain regions.

The Ising model has already been used in neuroscience to describe complex metastable states and dynamical processes using both anatomical networks (Das et al., 2014; Marinazzo et al., 2014) and functional connectivity networks extracted from fMRI analysis (Fraiman et al., 2009; Hudetz, Humphries, & Binder, 2014). However, these works used Monte Carlo algorithms to simulate the dynamics of the Ising model without particular justification. Neural message passing has been proposed as a common framework for many graph neural networks that can be used both for inference/predictions (Gilmer et al., 2017) and in machine learning applications (Lanchantin et al., 2019). In neuroscience, Friston and colleagues have already proposed the use of neuronal message-passing algorithms, such as BP and variational message passing, for modeling active inference, i.e. finding graphical and probabilistic models that explain the Bayesian functioning of the brain (Friston et al., 2017; Parr et al., 2019). In our work, we extend their intuition by connecting BP with a plausible model for neural dynamics and propose its use to explicitly model the transmission of information among macroscopic brain regions whose electrical activity is described by binary variables. In this sense, the direct application would be the prediction of functional activity given the structural underlying network. However, future studies should be devoted to exploring the possibility of estimating the underlying messages or structural connections that would fit experimental data, i.e. inverting the proposed model. Previous work has shown that this can indeed be possible for these algorithms under specific conditions (Dunn, Mørreaunet, & Roudi, 2015; Hertz et al., 2010; Ricci-Tersenghi, 2012; Roudi, Tyrcha, & Hertz, 2009).

Although the validation of this model is not straightforward -as there is no simple way to measure macroscopic electrical activities in all regions inside the brain, and the functional connectivity among them- we make here a preliminary validation by comparing the functional maps and correlation matrices obtained with the model to experimental maps corresponding to the well-known resting-state networks obtained from fMRI data (Beckmann et al., 2005).

The algorithms were naively implemented in Matlab and C++, and much optimization can be done, including the use of parallel or distributed computing. In our case, the BP converged in about 120 seconds, in average, while the SP took approximately $3 \times 10^4$ seconds. The computations were done in an Intel I3 processor, 2.20 GHz, and 8 GB of RAM.

### 4.2. The DMN is defined by the structural network in criticality

Our results showed that the fixed point of BP, when computed at criticality in the HC network and in the absence of external inputs (i.e. resting-state conditions), reproduces the main posterior activations of the experimental DMN map (**Fig. 2a** and **Fig. 2d**). The similarity between both maps supports a long-standing hypothesis suggesting that the appearance of the latter could be a product of the intrinsic properties of the anatomy of the brain (Deco & Jirsa, 2012; M. D. Greicius, Supekar, Menon, & Dougherty, 2009; Haimovici et al., 2013; Honey et al., 2009). In our case, this is a direct result of the BP algorithm that can be quantitatively compared with experimental maps, while even being simpler and more interpretable than previous works (Haimovici et al., 2013). The maps obtained for the random networks (null hypothesis networks), and the results of the ROC analysis, suggest that the appearance of the DMN map depends more strongly on the topology of the anatomical network (which are kept in RC2) than in the exact values of the connections. This has been important to find the similarity between the model and the experimental maps, as the exact values of the anatomical connectivity in HC can be derived from different tractography methodologies and even different measures of strength of connectivity (Iturria-Medina et al., 2007). This should be addressed in future studies by using larger databases for robust estimation of the structural network and by finding objective ways of thresholding the anatomical connectivity matrices. We would not discard the idea that the use of models like the one proposed here will help in deciding which are the relevant anatomical connections that lead to experimental functional maps, using machine learning strategies or inversion of the model.

In terms of a precise comparison between the areas appearing in the functional maps obtained with the BP, it is interesting to note that the model did not predict the activation of frontal areas. Although the functional roles of all regions pertaining to the DMN are still under debate, it has been proposed that these areas take part in a general ability to construct mental models of personally significant events, i.e. internal mentation. The posterior activations are related to memory retrieval and the frontal activations are more active in self-referential processing, reflecting other spontaneous cognitive processes in which the subject is inevitable involved when resting awake (Andrews-Hanna, 2012; Andrews-Hanna, Reidler, Huang, & Buckner, 2010; Buckner, 2012; Buckner et al., 2008). In that sense, the lack of frontal activations in the fixed-point map of the BP might support its relation to spontaneous cognitive processes which do not arise directly from the anatomical architecture of the brain. But it might be possible that this absence is related to the specific thresholding of the maps and the methodological issues mentioned above about the definition of the strength of anatomical connectivity. Determining which of the two explanations is more likely remains an open question for future studies.

Finally, it is important to emphasize that the BP maps predict the DMN only near the critical value of the control parameter of the network, which controls the global level of the strength of anatomical connectivity in the structural network. This supports the idea that the brain operates in a critical regime, according to the interpretation of its electrical activity as a product of the exchange of messages (synapses) between the macroscopic brain regions. This hypothesis has been largely proposed from different perspectives, from the direct anatomical network organization (Bullmore and Sporns 2009), to the power law of functional fMRI activations (Chialvo 2004; Chialvo 2010a), and to the strength of anatomical connection between macroscopic regions whose activation is modeled with a discrete state excitable dynamics following the Greenberg-Hastings model (Haimovici et al., 2013).

### 4.3. Functional resting-state connectivity from brain structure

Susceptibility Propagation (SP) computes the state correlation between every pair of nodes in a network. To our knowledge, this is the first time that it has been used to explore a dynamical model on a Human Connectome. In our case, these correlations can be interpreted as the zero-lag functional correlations among macroscopic regions of the brain. This magnitude has not been directly measured in neuroscience experiments since there are not non-invasive ways of measuring electrical activity in the whole brain. Actually, in the context of functional neuroimaging, functional connectivity is suggested to describe the relationship between the neuronal activation patterns of anatomically separated brain regions, reflecting the level of functional communication among regions. In particular, resting-state functional connectivity has been examined, for example, by measuring the level of temporal dependency among spontaneous functional MRI time-series, recorded during rest (Michael D Greicius et al., 2003). However, many methods have been used to quantitatively determine this temporal dependency and there is not yet a clear answer about which is the best stimulus-independent choice (Martijn P. van den Heuvel & Hulshoff Pol, 2010). In the case of electrophysiological data (EEG. MEG) this is still more challenging as measurements are gathered on or above the scalp, and one must firstly solve ill-posed inverse problems to determine the connectivity among brain regions (Sakkalis, 2011; van Diessen et al., 2015). Therefore, it is clear that the correlation matrix obtained with SP may not be directly comparable with the estimates coming from data-driven measures of temporal similarity. However, the modules (or subnetworks) derived from the correlation matrices obtained with SP can indeed be compared with the functional RSNs that have been consistently found in many fMRI studies, since the patterns obtained from clustering or finding modules of correlation matrices of fMRI time series are often quite similar to those obtained with ICA (Bellec, Rosa-Neto, Lyttelton, Benali, & Evans, 2010; Power et al., 2011; Sporns & Betzel, 2016; Thomas Yeo et al., 2011). Future work can also be oriented to study the dynamics of the BP and SP algorithms in order to derive predicted measures of temporal similarity of the activity of brain regions, which should be more directly evaluated with respect to experimental results.

In our results, the emergence of structured modularity was clear in the connectivity matrices obtained with the SP from the HC (**Fig. 5a**) and from the random network RC2 (**Fig. 5c**), -similar to the structure appearing directly from the structural connectivity (**Fig. 5d**)- and not in the case of the random network RC1 (**Fig. 5b**). This also points to the importance of a specific topology of the anatomical network for establishing a feasible organization of brain function. However, any structured organization in modules is not necessary indicative of the existence of proper activation patterns that underly actual brain processing. Therefore, more important than the pairwise comparison between the modular segregations of the four matrices is the analysis of its usefulness or validity as realistic functional connectivity matrices. To our knowledge, there are no direct measurements of the functional connectivity of macroscopic electrical activations. However, we could preliminarily explore the functional relevance of the different modules obtained in each case by comparing them with those experimental functional Resting State Networks (RSNs), which were translated to the Human Connectome of 998 regions in the work of Haimovici et al. 2013. The correspondence between modules and areas appearing in experimental resting-state networks (RSNs) was higher for the connectivity matrix obtained by the SP on the HC, which supports the idea that this model is able to predict the functional patterns resulting from real anatomical networks.

These results again suggest that the anatomical structure of the brain is crucial to understand the appearance of the different RSNs and they are consistent with an information flow defined by neural message passing between macroscopic regions of the brain. Moreover, it allows a new interpretation of the RSNs as the spontaneous correlations between brain regions that appear due to the anatomical connections among them and not necessarily because of a specific spontaneous dynamics of the activity of neuronal populations itself. This can help elucidating the question of the functional role of the RSNs, putting forward the hypothesis that these patterns are not formed for a specific function but just as the result of the natural activity and interaction in the absence of external stimulation. Nevertheless, given the indirect nature of the comparison between the correlation structure predicted by SP and the modules of experimental RSNs, further studies are needed to establish the physiological relevance of the connectivity values obtained with the model, as well as the influence of the clustering method used to find its modularity.

### 4.4. Predicting RSNs of pathological networks

This paper presents a new model for predicting functional activity/connectivity from the underlying anatomical network in the case of resting-state conditions. Obviously, a wider usefulness of the model would derive from demonstrating its ability to predict functional activity/connectivity also in task-related conditions. A more straightforward, but not less important, application is also to explore whether the model is able to predict the changes in functional activity/connectivity related to changes in the anatomical network. Here we presented a preliminary exploration of this question by following the recent literature on the alterations found in anatomical networks and the resting-state functional connectivity in two well-studied pathologies.

Firstly, we simulated a structural network in a patient with Alzheimer's Disease (AD), which has been found to present a general degeneration of fiber tracts in almost all the brain, with implications on the decrease of interactions between neuronal masses (Griffa et al., 2013). Then, our altered network was created by multiplying all connection weights of the HC by 0.8. The fixed-point solution of the BP on this network, using the critical value of the control parameter obtained from the original HC, led to weaker activations, where the structure of the back hub of the DMN disappeared (**Fig. 9a**). This result is in agreement with studies on the DMN in AD patients, which entrust the devastating effects of this disease on higher cognitive functions (including the ability to remember and imagine) to the preferential disruption of expanded brain systems that are important to the default mode network (Buckner, 2012). When using the critical parameter computed for the altered network, the DMN was identical to the one predicted from the original HC, which follows directly from the global scaling effect of the critical parameter, such that recomputing the critical value will take the scaling back to the original conditions. This suggests the interpretation that, in an AD structural network, the brain is functioning in super-critical conditions (lower connection weights), which supports the hypothesis and findings about the loss of emergent properties of complex networks (Griffa et al., 2013).

In short, the long-term correlation matrix obtained by SP in the AD network showed a decrease in the functional connectivity, which agrees with experimental studies (M. D. Greicius, Srivastava, Reiss, & Menon, 2004; Supekar, Menon, Rubin, Musen, & Greicius, 2008; Wang et al., 2007; Wu et al., 2011). The cluster organization of such matrix was more similar to the direct clustering of the anatomical HC matrix than to the cluster division of the correlation matrix obtained with SP from the HC (**Fig. 10c** and **Fig. 6b**). This is consistent with the idea of working in a super-critical condition where only anatomical connections matters and there is not functional exchange. However, when compared to experimental resting-state networks (RSNs), the modules from SP-AD showed higher values of evaluation metrics than the modules from the anatomical HC matrix (**Fig. 7b**), which suggests that there the functional interactions are not completely absent. Other ways of simulating AD-related structural networks, would be useful to validate if this model, and its interpretation, are able to account for experimental results.

Secondly, we examined the conditions of anomalies in the Corpus Callosum (CC), which comprises the principal interhemispheric white matter tract carrying nervous fibers that connect neuronal populations located in the two cerebral hemispheres. Anomalies can be congenital, -being agenesis (i.e. total absence of the corpus callosum) and hypogenesis (also called "partial agenesis") the most important ones-, and acquired due to diseases or conditions (aggressive tumors, demyelination, traumatic brain injury) or due to the surgical procedure known as Callosotomy (Ho, Moonis, Ginat, & Eisenberg, 2013). Here we studied the condition in which all connections between both hemispheres are impaired, which correspond to Callosotomy or complete agenesis. Despite the obvious effect on the functional interaction of neuronal populations, the lesions of the CC usually lead to mechanical and other changes in the brain that are not necessarily related to functional activity, but can imply local impairments of the normal functional behavior of neuronal masses.

In this context, several studies have shown that there are changes in the functional activity/connectivity in patients with these lesions, particularly in the DMN and RSNs (Peng & Hsin, 2017). We found, in accordance with those studies, that there is indeed a disruption of the DMN for this network as estimated by BP using the critical parameter of the normal HC (**Fig. 9b**). However, if the control parameter is changed to the critical value obtained for the altered CC network, the DMN map given by BP is again much similar to the DMN found in healthy patients. Surprisingly, normal DMN maps have been reported in resting-state studies in subjects with agenesis and in follow-up studies on patients that underwent Callosotomy several years ago (Roland et al., 2017; Tyszka et al., 2012). Applying our model interpretation, we should then propose that the criticality of the network can be restored in the long term even for compensating such a large damage in the structural topology. As expected, the correlation matrix obtained with SP from the CC network showed only nonzero values for connections within each hemisphere, but keeping a spatially clustered organization. The clustering was not completely similar to others obtained from SP-HC or from the anatomical HC matrix, but it led to modules that had a high correspondence (within hemispheres) with experimental RSNs. Noteworthy the SP-CC clustering showed a similarly high symmetry than that of the SP-HC. Whether or not these aspects have a real relevance for brain function organization, remains as future directions of research.

Despite the similarity of our model's predictions of activity/connectivity in simulated pathological networks, with current experimental and clinical studies using anatomical and functional data, both cases analyzed here have offered interesting clues on the role of criticality in the emergence of functional patterns in the brain. Criticality has been claimed for explaining the brain since long ago (Bak, 1996), but it is in the last decade that a group of research articles have presented evidence of its presence and discussing its implications (Deco & Jirsa, 2012; Fraiman et al., 2009; Haimovici et al., 2013; Hudetz et al., 2014; Marinazzo et al., 2014; Stramaglia, 2014; Tagliazucchi et al., 2012). However, criticality does not mean the same thing when studying the small-world properties of anatomical or functional (mathematically derived) networks (Bullmore & Sporns, 2009; Gafarov, 2016; Rubinov & Sporns, 2010; M.P. van den Heuvel, Stam, Boersma, & Hulshoff Pol, 2008), or the power-law distribution of functional data (Chialvo, 2004, 2010; Expert et al., 2011; Fraiman et al., 2009; Fraiman, Chialvo, & Breakspear, 2012; Sporns, Chialvo, Kaiser, & Hilgetag, 2004; Tagliazucchi et al., 2012), or the information transfer in an Ising model (Marinazzo et al., 2014) or when proposing non-linear functional dynamics on anatomically connected nodes (Haimovici et al., 2013). If it may well be that all these criticalities are needed for the emergence of complex brain functioning, we would like to stress that in this work we are dealing with a functional criticality, which fortunately depends on just one parameter, that is found by a statistical-physics simulation of activity within a given structural network. In other words, the critical behavior in BP and SP is not related to the network topology alone, but it is also determined by the model assumed for functional interactions and the nature of possible external influences.

In this context, it is interesting to note that some of our results suggest that the same functional maps can emerge from anatomical networks with fundamentally different topologies (not just with the allegedly small inter-subject variations). Even more, the functional maps can change with time from pathological patterns to normal patterns in the same structural network, which indicates that there might be functional mechanisms involved in reestablishing the critical conditions. Some recent studies have shown that the phenomenon of nerve growth can be driven by functional activity, which means that the anatomical reorganization is oriented to functional optimality (Gafarov, 2016, 2018). However, in the case of severe damage or change in the topology of brain network, such as Callosotomy in adults, this mechanism does not seem to account for the full restoration of criticality. According to the interpretation provided by our model, criticality might arise just from adjusting the general strength of interactions, but can also be mediated by the use of a message-passing model of the interactions. In this sense, the BP and SP can be seen as natural choices for modeling the mechanism of moving "towards criticality", as the message-passing strategy is closer to the actual interaction process among neuronal populations. Indeed, the view of BP as a process of minimizing the free energy of a system, has advanced the idea that this type of message-passing algorithms entail self-organized criticality as an intrinsic emergent property, and discuss its implications in the context of effective connectivity and neuroimaging (Friston, Kahan, Razi, Stephan, & Sporns, 2014; Friston et al., 2017).

## 5. Conclusions

In this work we modelled macroscopic brain electrical activity as a collection of binary variables placed in the nodes of the Human Connectome (HC) and interacting cooperatively with a strength defined by the corresponding anatomical connectivity matrix. The dynamics of the system was defined through neural message-passing algorithms near the critical point of the model. The global attractor of this dynamics resembles the posterior hub of the Default Mode Network (DMN), which implies that the DMN can be interpreted as an activation pattern which result from the intrinsic properties of the cerebral anatomical network at rest. Future studies are needed to determine if the frontal hub reflects other spontaneous cognitive processes or if it is a flaw of the model. On the other hand, the correlation between the activity of regions defined within this dynamics was estimated using Susceptibility Propagation. The modules obtained from the clustering of this correlation matrix, resembled many of the experimental Resting State Networks (RSN), again suggesting that the latter result from spontaneous long-range electrical interactions in the brain. Although further validation is needed, this model offers a physiologically plausible way to obtain and interpret the functional maps and the functional connectivity matrices in resting-state. It may also allow the prediction of changes in these functional maps/networks induced by changes in the anatomical connectome, including the modelling of neuro-development and aging-related pathologies. In a preliminary simulation of AD and lesions of the Corpus Callosum, the model was able to show the disruption of the DMN but keeping a modular structure of the activity in coherence resting-state networks, as has been found in the research and clinical practice. Criticality of the anatomical network and of the neural message-passing mechanisms seem to be an important factor for the emergence of realistic patterns of activations and correlations in all cases, as the model also primed an explanation to the phenomena of long-term functional reorganization.

We believe that this model is a promising tool for exploring the relationship between the anatomical substrate and the function of the brain. The use of neural message-passing algorithms offers a way to model actual mechanisms of neuronal interaction and also opens the door to machine learning applications. Moreover, it could have implications on the interpretation of brain functioning within a general framework of universal laws of physics based on self-organized criticality. Future development of the model should consider including the effects of external/internal stimulation of particular regions to model other brain states and more complex cognitive processes. In addition, finding reliable relations between the model output and real experimental data should pave the way to address the inversion of the model, such that the changes in the structural connectivity underlying measured functional data can be estimated. Finally, an especially interesting direction for future research would be the possibility of adapting the message-passing strategy to the study of dynamic connectivity through dynamic graph models of brain networks (Khambhati, Sizemore, Betzel, & Bassett, 2018).

## Conflict of Interest

The authors declare that the research was conducted in the absence of any commercial or financial relationships that could be construed as a potential conflict of interest.

## Funding

This work has received partial funding from the European Union Horizon 2020 research and innovation program MSCA-RISE-2016 under grant agreement No 734439 INFERNET. Support was also received by CONICYT program MEC 80170124.

## Author Contributions

JPG contributed to developing the theory and its implementation, performing the analysis and writing the paper.

EMM contributed to the design of the study, the interpretation of results, and writing of the paper.

EA developed visualization tools.

PVH participated in the interpretation of results.

RM contributed to the design of the study, developing of theory, the interpretation of results and writing of the paper.

## Acknowledgments

We thank Prof. Olaf Sporns and Dr. Daniele Marinnazo for sharing the structural connectome and Prof. Dante Chialvo who provided the mask of experimental RSNs. Authors would like to especially acknowledge the collaboration in this work from Alejandro Lage for interesting discussions and helpful suggestions.

## Supplementary Material

### A. Expression for the magnetization in the Belief Propagation (BP) algorithm

In this work, we have applied the BP algorithm on an Ising model of brain macroscopic regions that correspond to the nodes of the network, to compute the functional patterns of brain electrical activation. In this section we give the mathematical details of the algorithm using the original terminology of statistical physics where nodes correspond to spins, and the output of the algorithm (the brain functional activation) corresponds to the local magnetization of the system of spins. According to the BP the marginal distributions of probability $p_i(s_i)$ are roughly calculated from a variable called belief, and we denote the belief of node $i$ as $b_i(s_i)$ which depends on the state of the spin (Yedidia et al., 2003). The $b_i(s_i)$ represents the probability that the node $i$ is in the state $s_i$. It depends on an external local field on it and the information it receives from its first neighbors $j$ in the form of messages $m_{j\to i} = m_{ji}$[4], (**Fig. A.1**).

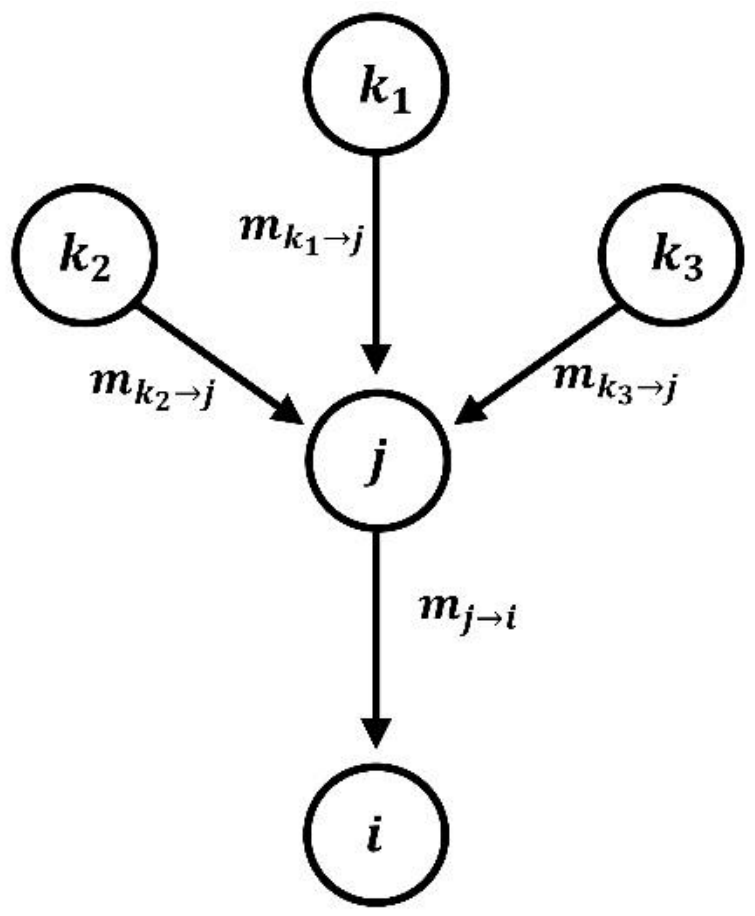

**Fig A.1.** Messages that enter node j to calculate the message that node j sends to node i.

This message represents the information sent by node $j$ about which state is most favorable for node $i$, where its equation for an Markov Random Field (MRF) with the Ising model is (Yedidia et al., 2003):

$$m_{ji}^{t_d+1}(s_i) = k\sum_{s_j} e^{\beta\left(J_{ji}s_j s_i + h_j s_j\right)} \prod_{l\in N(j)\backslash i} m_{lj}^{t_d}(s_j) \tag{A.1}$$

Where $N(j)\backslash i$ represents the set of neighbors of node $j$ except $i$; $k$ is a normalization constant determined by: $m_{ji}^{t_d}(1) + m_{ji}^{t_d}(-1) = 1$; the term $e^{\beta h_j s_j}$ is the local contribution of node $j$; $\beta$ is the control parameter (inverse of

[4] For convenience in this section we use the $m_{ji}$ instead of $m_{j\to i}$, which represents exactly the same magnitude. This variable $m_{ji}$ with two sub-indices represents the messages, while the variable $m_i$ with only one sub-index represents the magnetization (functional activation) of the node $i$.

*T*); $h_j$ is an external field in *j*; $s_j$ are the different states of the spin $s_j = \pm 1$ (that correspond to the brain region being active or non-active in our Ising model); the term $e^{\beta J_{ji} s_j s_i}$ describes the interaction of neighboring nodes *j* and *i*; and $t_d$: refers to the discrete time (iterations).

The algorithm consists of iterating Eq. (A.1) over each message of the network and waiting for the message set to converge to a fixed point. The condition of convergence is given by the variation of the message:

$$\left|\Delta d\right|_{\max} = \left|m_{ji}^{old} - m_{ji}^{new}\right|_{\max} \leq \delta_{\max} \quad with \;\; \delta_{\max} = 10^{-6} \;\; \forall i \tag{A.2}$$

where the sub-index *max* in the relative variation refers to the greatest variation obtained during a cycle of iterations, that is, over updated messages for all nodes in the network.

Once the messages converged, we can compute the belief of each node *i*, which is proportional to the product of the local contribution of the node ($e^{\beta h_i s_i}$) and all the messages sent by neighbors $N(i)$ of node *i* (**Fig. A.2**) (Yedidia et al., 2003):

$$b_i\left(s_i\right) = k e^{\beta h_i s_i} \prod_{j \in N(i)} m_{j \to i}\left(s_i\right) \tag{A.3}$$

The proportionality constant *k* is determined by the normalization condition:

$$\sum_{\{s_i\}} b_i\left(s_i\right) = 1 \quad \to \quad k = \frac{1}{e^{\beta h_i} \prod_{j \in N(i)} m_{j \to i}(+1) + e^{-\beta h_i} \prod_{j \in N(i)} m_{j \to i}(-1)} \tag{A.4}$$

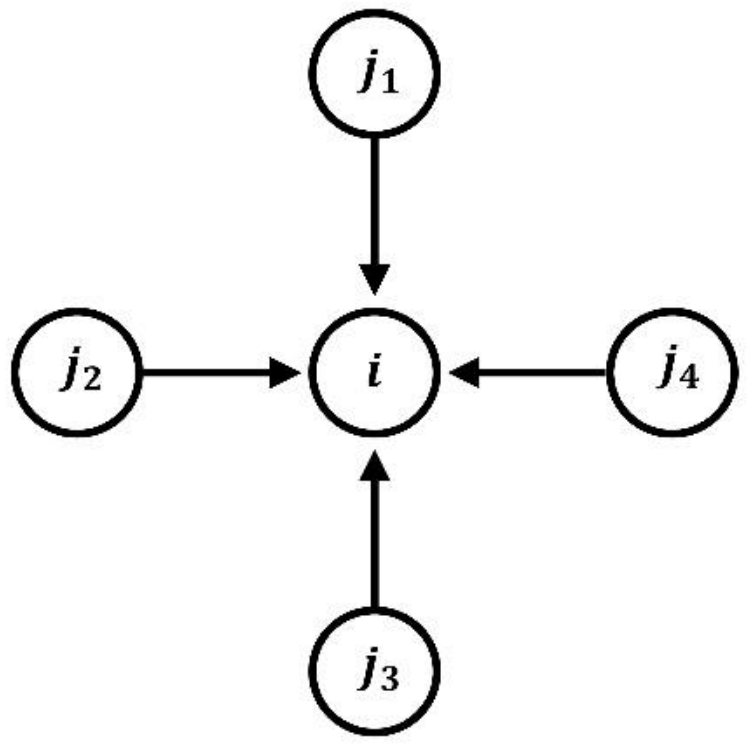

**Fig A.2.** Messages from the neighbors of node i to calculate the belief $b_i$.

Then, given that the belief represents the probability distribution, the local magnetization of the spin is calculated by:

$$m_i = \langle s_i \rangle = \sum_{s_i} s_i p_i(s_i) = \sum_{s_i} s_i b_i(s_i) \quad (A.5)$$

which yields:

$$\boxed{m_i = b_i(+1) - b_i(-1)} \quad (A.6)$$

## B. Pseudo code for the BP algorithm

***ALGORITHM B.1. BELIEF PROPAGATION.***

| | |
|---|---|
| **STEP 1:** | Initialize all messages randomly $m_{ij}$ |
| **STEP 2:** | **for** $i = 0$ to IMAX do // IMAX is the maximum number of iterations (**BP**) |
| **STEP 3:** | The messages corresponding to the BP given by Eq. (A.1) are updated. |
| **STEP 4:** | **if** convergence condition holds, **then** |
| **STEP 5:** | $i =$ IMAX // stop the iteration |
| **STEP 6:** | e**nd if** |
| **STEP 7:** | **end for** |
| **STEP 8:** | Local activations are calculated with Eq. (A.6). |

## C. Exact relationship between the BP equations and the neurodynamical equation of a Hopfield continuous network

The equation that describes the dynamics of a continuous network of Hopfield is:

$$\frac{dv_i(t)}{dt} = -v_i(t) + f\left(\sum_l w_{li} v_l(t)\right) + \beta K_i(t) \quad (C.1)$$

Where $v_i$ describes the activity of the neuron *i* (this original terminology indeed refers to large neuronal masses); $f(x)$ is the activation function (typically non-linear, e.g. sigmoid function); $w_{li}$ is the connection or synaptic weight, which is positive and symmetrical for the Hopfield model ($w_{li} = w_{il},\ \forall\ l, i$); and $K_i(t)$ represents the external noise.

In order to establish the exact relationship between Hopfield and BP, the translation of the Markov Random Field to a Hopfield network must be carried out by implementing the following three steps (Ott & Stoop, 2006):

1. Reduction in the number of messages per connection through the re-parameterization:

$$\operatorname{Tanh} n_{ji} = m_{ji}(+1) - m_{ji}(-1) \tag{C.2}$$

Because of this, the updating rule Eq. (A.1) is transformed as a function of the $n_{ji}$ in the following way:

$$\tanh n_{ji}^{t_d+1} = m_{ji}^{t_d+1}(+1) - m_{ji}^{t_d+1}(-1) = k\left[\sum_{s_j} e^{\beta(J_{ji}s_j + h_j s_j)} \prod_{l\in N(j)\setminus i} m_{lj}^{t_d}(s_j) - \sum_{s_j} e^{\beta(-J_{ji}s_j + h_j s_j)} \prod_{l\in N(j)\setminus i} m_{lj}^{t_d}(s_j)\right] =$$

$$= \frac{e^{\beta(J_{ji}+h_j)} \prod_{l\in N(j)\setminus i} m_{lj}^{t_d}(+1) + e^{-\beta(J_{ji}+h_j)} \prod_{l\in N(j)\setminus i} m_{lj}^{t_d}(-1) - e^{-\beta(J_{ji}-h_j)} \prod_{l\in N(j)\setminus i} m_{lj}^{t_d}(+1) - e^{\beta(J_{ji}-h_j)} \prod_{l\in N(j)\setminus i} m_{lj}^{t_d}(-1)}{e^{\beta(J_{ji}+h_j)} \prod_{l\in N(j)\setminus i} m_{lj}^{t_d}(+1) + e^{-\beta(J_{ji}+h_j)} \prod_{l\in N(j)\setminus i} m_{lj}^{t_d}(-1) + e^{-\beta(J_{ji}-h_j)} \prod_{l\in N(j)\setminus i} m_{lj}^{t_d}(+1) + e^{\beta(J_{ji}-h_j)} \prod_{l\in N(j)\setminus i} m_{lj}^{t_d}(-1)} =$$

$$= \frac{\left[e^{\beta J_{ji}} - e^{-\beta J_{ji}}\right] e^{\beta h_j} e^{\sum_{l\in N(j)\setminus i} \log m_{lj}^{t_d}(+1)} - \left[e^{\beta J_{ji}} - e^{-\beta J_{ji}}\right] e^{-\beta h_j} e^{\sum_{l\in N(j)\setminus i} \log m_{lj}^{t_d}(-1)}}{\left[e^{\beta J_{ji}} + e^{-\beta J_{ji}}\right] e^{\beta h_j} e^{\sum_{l\in N(j)\setminus i} \log m_{lj}^{t_d}(+1)} + \left[e^{\beta J_{ji}} + e^{-\beta J_{ji}}\right] e^{-\beta h_j} e^{\sum_{l\in N(j)\setminus i} \log m_{lj}^{t_d}(-1)}} =$$

$$= \frac{\left[e^{\beta J_{ji}} - e^{-\beta J_{ji}}\right]\left[e^{\beta h_j + \sum_{l\in N(j)\setminus i} \log m_{lj}^{t_d}(+1)} - e^{-\beta h_j + \sum_{l\in N(j)\setminus i} \log m_{lj}^{t_d}(-1)}\right]}{\left[e^{\beta J_{ji}} + e^{-\beta J_{ji}}\right]\left[e^{\beta h_j + \sum_{l\in N(j)\setminus i} \log m_{lj}^{t_d}(+1)} + e^{-\beta h_j + \sum_{l\in N(j)\setminus i} \log m_{lj}^{t_d}(-1)}\right]}$$

Using the definition of the hyperbolic tangential: $\tanh x = e^x - e^{-x}/e^x + e^{-x}$:

$$\tanh n_{ji}^{t_d+1} = \tanh(\beta J_{ji}) \frac{\left[e^{\beta h_j + \sum_{l\in N(j)\setminus i} \log m_{lj}^{t_d}(+1)} - e^{-\left(\beta h_j + \sum_{l\in N(j)\setminus i} \log \frac{1}{m_{lj}^{t_d}(-1)}\right)}\right]}{\left[e^{\beta h_j + \sum_{l\in N(j)\setminus i} \log m_{lj}^{t_d}(+1)} + e^{-\left(\beta h_j + \sum_{l\in N(j)\setminus i} \log \frac{1}{m_{lj}^{t_d}(-1)}\right)}\right]} = \tanh(\beta J_{ji}) \frac{\left[e^{\beta h_j + \sum_{l\in N(j)\setminus i} \frac{1}{2}\log \frac{m_{lj}^{t_d}(+1)}{m_{lj}^{t_d}(-1)}} - e^{-\left(\beta h_j + \sum_{l\in N(j)\setminus i} \frac{1}{2}\log \frac{m_{lj}^{t_d}(+1)}{m_{lj}^{t_d}(-1)}\right)}\right]}{\left[e^{\beta h_j + \sum_{l\in N(j)\setminus i} \frac{1}{2}\log \frac{m_{lj}^{t_d}(+1)}{m_{lj}^{t_d}(-1)}} + e^{-\left(\beta h_j + \sum_{l\in N(j)\setminus i} \frac{1}{2}\log \frac{m_{lj}^{t_d}(+1)}{m_{lj}^{t_d}(-1)}\right)}\right]}$$

$$= \tanh(\beta J_{ji}) \tanh\left[\beta h_j + \sum_{l\in N(j)\setminus i} \frac{1}{2} \log \frac{m_{lj}^{t_d}(+1)}{m_{lj}^{t_d}(-1)}\right]$$

Using the parametrization Eq. (C.2) and the normalization condition $m_{j\to i}(+1) + m_{j\to i}(-1) = 1$:

$$m_{j\to i}(+1)=\frac{1}{2}\left(1+\tanh n_{j\to i}\right) \qquad m_{j\to i}(-1)=\frac{1}{2}\left(1-\tanh n_{j\to i}\right) \tag{C.3}$$

We get:

$$\tanh n_{ji}^{t_d+1}=\tanh\left(\beta J_{ji}\right)\tanh\left\{\beta h_i+\sum_{j\in N(i)}\frac{1}{2}\log\left[\frac{1+\tanh n_{j\to i}^{t_d}}{1-\tanh n_{j\to i}^{t_d}}\right]\right\}$$

Then, calling $U=\tanh n_{j\to i}^{t_d}$ and using the relationship:

$$U=\tanh\left\{\frac{1}{2}\log\left[\frac{1+U}{1-U}\right]\right\} \tag{C.4}$$

We finally get:

$$n_{ji}^{t_d+1}=\tanh^{-1}\left\{\tanh\left(\beta J_{ji}\right)\tanh\left[\beta h_j+\sum_{l\in N(j)\setminus i}n_{lj}^{t_d}\right]\right\} \tag{C.5}$$

Therefore, for each connection $j\to i$ there is only one message. Consequently, we can write the local magnetizations Eq. (A.6) in terms of the new messages $n_{ji}$.

We use the relationships Eq. (C.4):

$$m_i=b_i(+1)-b_i(-1)=\frac{b_i(+1)-b_i(-1)}{b_i(+1)+b_i(-1)}=\tanh\left\{\frac{1}{2}\log\left[\frac{1+m_i}{1-m_i}\right]\right\}=\tanh\left\{\frac{1}{2}\log\left[\frac{1+\frac{b_i(+1)-b_i(-1)}{b_i(+1)+b_i(-1)}}{1-\frac{b_i(+1)-b_i(-1)}{b_i(+1)+b_i(-1)}}\right]\right\}$$

$$=\tanh\left\{\frac{1}{2}\log\left[\frac{b_i(+1)}{b_i(-1)}\right]\right\}$$

Taking the definition of $b_i$ Eq. (A.3):

$$m_i=\tanh\left\{\frac{1}{2}\log\left[\frac{e^{\beta h_i}\prod_{j\in N(i)}m_{j\to i}(+1)}{e^{-\beta h_i}\prod_{j\in N(i)}m_{j\to i}(-1)}\right]\right\}=\tanh\left\{\frac{1}{2}2\beta h_i+\frac{1}{2}\left[\sum_{j\in N(i)}\log m_{j\to i}(+1)-\sum_{j\in N(i)}\log m_{j\to i}(-1)\right]\right\}$$

$$=\tanh\left\{\beta h_i+\sum_{j\in N(i)}\frac{1}{2}\log\frac{m_{j\to i}(+1)}{m_{j\to i}(-1)}\right\}$$

Using the transformations Eq. (C.3):

$$m_i = \tanh\left\{\beta h_i + \sum_{j\in N(i)} \frac{1}{2}\log\left[\frac{1+\tanh n_{j\to i}}{1-\tanh n_{j\to i}}\right]\right\}$$

Finally, using again the identity Eq. (C.4) we obtain:

$$\boxed{m_i = \tanh\left(\beta h_i + \sum_{l\in N(i)} n_{li}\right)} \tag{C.6}$$

2. Translation of the discrete system towards the continuous system over time:

There are many ways to make this translation, but the most elegant is the one derived from the definition of the derivative:

$$\frac{dn_{ji}(t)}{dt} \stackrel{def}{=} \lim_{\tau\to 0}\frac{n_{ji}(t+\tau)-n_{ji}(t)}{\tau} \tag{C.7}$$

Where it is assumed that now the discrete time $t_d$ is continuous $t$, and the instant $t+\tau$ corresponds to $t_d+1$ (Akkermans, 1992); therefore, Eq. (C.7) results in:

$$\frac{dn_{ji}(t)}{dt} \stackrel{def}{=} n_{ji}(t_d+1)-n_{ji}(t_d) = -n_{ji}^{t_d} + n_{ji}^{t_d+1} \tag{C.8}$$

Eq. (C.5) results in:

$$\frac{dn_{ji}(t)}{dt} = -n_{ji}(t) + \tanh^{-1}\left\{\tanh\left(\beta J_{ji}\right)\tanh\left[\beta h_j + \sum_{l\in N(j)\setminus i} n_{lj}(t)\right]\right\} \tag{C.9}$$

The stability and convergence at the same fixed point for both Eq. (C.5) and Eq. (C.9), is guaranteed (Ott & Stoop, 2006).

3. Translation of Eq. (C.9) into the Hopfield network equation:

For comparing Eq. (C.1) and Eq. (C.9), we realize that, under certain conditions, the $n_{ji}$ can be identified with $w_{ji}v_j$, i.e. the messages represent the contribution of activity of the neighbor neuron $j$ on $i$: $n_{ji} = w_{ji}v_j^i$. Rewriting then Eq. (C.9) we get:

$$\frac{dw_{ji}v_j^i}{dt} = -w_{ji}v_j^i + \tanh^{-1}\left\{ w_{ji} \tanh\left[ \beta h_j + \sum_{l\in N(j)} w_{lj}v_l^j - w_{ij}v_i^j \right] \right\}, \tag{C.10}$$

Where connection weights are defined as $w_{ji} = \tanh(\beta J_{ji})$. Considering these weights to be relatively weak, $w_{ji} \ll 1$, the approximation $\tanh^{-1}(x) \approx x$, holds. Also, if each neuron receives a large number of inputs then the simple contribution $w_{ij}v_i^j$ can be neglected. Therefore, Eq. (C.10) is simplified to:

$$\frac{dw_{ji}v_j^i}{dt} = -w_{ji}v_j^i + w_{ji} \tanh\left[ \beta h_j + \sum_{l\in N(j)} w_{lj}v_l^j \right] \tag{C.11}$$

Simplifying the terms $w_{ji}$, we get:

$$\frac{dv_j^i}{dt} = -v_j^i + \tanh\left[ \beta h_j + \sum_{l\in N(j)} w_{lj}v_l^j \right] \tag{C.12}$$

Where the uniform initial conditions $v_j^1(0) = v_j^2(0) = \cdots = v_j^{c_j}(0)$ preserve their uniformity over time for all $j$, that is, $v_j^1(t) = v_j^2(t) = \cdots = v_j^{c_j}(t)$. This means that the subset $v_j^1 = v_j^2 = \cdots = v_j^{c_j}$ is invariant to the dynamics of Eq. (C.12). Therefore, for each $j$ we can replace $v_j^i$ with $v_j$, and rewrite the final equation for every neuron as:

$$\frac{dv_i}{dt} = -v_i + \tanh\left[ \beta h_i + \sum_{l\in N(i)} w_{li}v_l \right] \tag{C.13}$$

Using $\tanh(x+y) \approx \tanh(x) + \tanh(y)$ if $y \ll 1$, and with $y = \beta h_i$, Eq. (C.13) matches with the equation of the Hopfield model Eq. (C.1). After convergence to the attractor fixed point, the local magnetization simply corresponds to the activity of the neuron:

$$\tanh\left[ \beta h_j + \sum_{l\in N(j)} w_{lj}v_l^j \right] \approx \tanh\left( \sum_l w_{li}v_l(t) \right) + \beta K_i(t) \quad \Rightarrow \quad v_i(t=\infty) = m_i \tag{C.14}$$

## D. Expressions for the local susceptibility in the Susceptibility Propagation (SP) algorithm

In this section we first need to introduce a convenient notation for Belief Propagation. The messages already defined in Eq. (A.1), can be written as $m_{j\to i}(s_i) = q_{j\to i}(s_i)$ (unlike Section A, we use the notation $m_{j\to i}$ to improve the understanding of the expressions) (Charles Ollion, 2010):

$$q_{j\to i}\left(s_i\right)=k\sum_{s_j}e^{\beta J_{ji}s_js_i}p_{j\to i}\left(s_j\right) \tag{D.1}$$

Where the messages $p_{j\to i}$ include the factors corresponding to the external field, according to:

$$p_{j\to i}\left(s_j\right)=e^{\beta h_j s_j}\prod_{l\in N(j)\backslash i}q_{l\to j}\left(s_j\right) \tag{D.2}$$

Finally, the belief $b_i(s_i)$ results in:

$$p_i\left(s_i\right)=ke^{\beta h_i s_i}\prod_{l\in N(i)}q_{l\to i}\left(s_i\right) \tag{D.3}$$

And from (A.4):

$$k=\frac{1}{e^{\beta h_i}\prod_{l\in N(i)}q_{l\to i}\left(+1\right)+e^{-\beta h_i}\prod_{l\in N(i)}q_{l\to i}\left(-1\right)} \tag{D.4}$$

A convenient way to use Eq. (D.1) and Eq. (D.2) is known as the log-likelihood notation:

$$h_{i\to j}=\frac{1}{2}\log\frac{p_{i\to j}\left(+1\right)}{p_{i\to j}\left(-1\right)} \tag{D.5}$$

$$\begin{aligned}h_{i\to j}&=\frac{1}{2}\log\frac{e^{\beta h_i}\prod_{l\in N(i)\backslash j}q_{l\to i}\left(+1\right)}{e^{-\beta h_i}\prod_{l\in N(i)\backslash j}q_{l\to i}\left(-1\right)}=\frac{1}{2}2\beta h_i+\frac{1}{2}\left(\sum_{l\in N(i)\backslash j}\log q_{l\to i}\left(+1\right)-\sum_{l\in N(i)\backslash j}\log q_{l\to i}\left(-1\right)\right)\\&=\beta h_i+\sum_{l\in N(i)\backslash j}\frac{1}{2}\log\frac{q_{l\to i}\left(+1\right)}{q_{l\to i}\left(-1\right)}\end{aligned}$$

$$u_{i\to j}=\frac{1}{2}\log\frac{q_{i\to j}\left(+1\right)}{q_{i\to j}\left(-1\right)} \tag{D.6}$$

$$u_{i\to j}=\frac{1}{2}\log\frac{e^{\beta J_{ji}}p_{i\to j}\left(+1\right)+e^{-\beta J_{ji}}p_{i\to j}\left(+1\right)}{e^{-\beta J_{ji}}p_{i\to j}\left(-1\right)+e^{\beta J_{ji}}p_{i\to j}\left(-1\right)}$$

Making $U = \frac{e^{\beta J_{ji}} p_{i\to j}(+1) + e^{-\beta J_{ji}} p_{i\to j}(-1) - e^{-\beta J_{ji}} p_{i\to j}(+1) - e^{\beta J_{ji}} p_{i\to j}(-1)}{e^{\beta J_{ji}} p_{i\to j}(+1) + e^{-\beta J_{ji}} p_{i\to j}(-1) + e^{-\beta J_{ji}} p_{i\to j}(+1) + e^{\beta J_{ji}} p_{i\to j}(-1)}$, we have from the identity (C.4) that:

$$u_{i\to j} = \frac{1}{2}\log\frac{1+U}{1-U} = \tanh^{-1} U$$

$$\tanh u_{i\to} = U = \frac{(e^{\beta J_{ij}} - e^{-\beta J_{ij}})}{(e^{\beta J_{ij}} + e^{-\beta J_{ij}})}\frac{(p_{i\to j}(+1) - p_{i\to j}(-1))}{(p_{i\to j}(+1) + p_{i\to j}(-1))} = \tanh\left(\beta J_{ji}\right)\frac{\sqrt{\frac{p_{i\to j}(+1)}{p_{i\to j}(-1)}}\left(\sqrt{\frac{p_{i\to j}(+1)}{p_{i\to j}(-1)}} - \sqrt{\frac{p_{i\to j}(-1)}{p_{i\to j}(+1)}}\right)}{\sqrt{\frac{p_{i\to j}(+1)}{p_{i\to j}(-1)}}\left(\sqrt{\frac{p_{i\to j}(+1)}{p_{i\to j}(-1)}} + \sqrt{\frac{p_{i\to j}(-1)}{p_{i\to j}(+1)}}\right)}$$

$$= \tanh\left(\beta J_{ji}\right)\tanh\left(h_{i\to j}\right)$$

Which results in simple notation for new messages to be updated with BP:

$$\boxed{h_{i\to j} = \beta h_i + \sum_{l\in N(i)\setminus j} u_{l\to i}} \tag{D.7}$$

$$\boxed{u_{i\to j} = \tanh^{-1}\left[\tanh\left(\beta J_{ji}\right)\tanh\left(h_{i\to j}\right)\right]} \tag{D.8}$$

Then, similarly to the deduction in Eq. (A.5) and now taking the probability function from Eq. (D.3) with the normalization constant Eq. (D.4), we have:

$$m_i = \langle s_i \rangle = \sum_{s_i} s_i p_i(s_i) = p_i(+1) - p_i(-1) \tag{D.9}$$

Using the identity Eq. (C.4):

$$m_i = \frac{p_i(+1) - p_i(-1)}{p_i(+1) + p_i(-1)} = \tanh\frac{1}{2}\log\left[\frac{1+m_i}{1-m_i}\right] = \tanh\frac{1}{2}\log\left[\frac{1+\frac{p_i(+1)-p_i(-1)}{p_i(+1)+p_i(-1)}}{1-\frac{p_i(+1)-p_i(-1)}{p_i(+1)+p_i(-1)}}\right] = \tanh\frac{1}{2}\log\left[\frac{p_i(+1)}{p_i(-1)}\right]$$

$$= \tanh\left\{\frac{1}{2}\log\left[\frac{e^{\beta h_i}\prod_{l\in N(i)\setminus j} q_{l\to i}(+1)}{e^{-\beta h_i}\prod_{l\in N(i)\setminus j} q_{l\to i}(-1)}\right]\right\} = \tanh\left\{\frac{1}{2}2\beta h_i + \frac{1}{2}\left(\sum_{l\in N(i)\setminus j}\log q_{l\to i}(+1) - \sum_{l\in N(i)\setminus j}\log q_{l\to i}(-1)\right)\right\}$$

$$= \tanh\left\{\beta h_i + \sum_{l\in N(i)\setminus j}\frac{1}{2}\log\frac{q_{l\to i}(+1)}{q_{l\to i}(-1)}\right\}$$

The resulting expression for $m_i$ is:

$$\boxed{m_i = \tanh\left(\beta h_i + \sum_{l \in N(i)} u_{l \to i}\right)} \tag{D.10}$$

In statistical mechanics, the correlation function $\pi_{ij}$ between the states of pairs of nodes (*i* and *j*) at distant locations, is defined as the thermal average (canonical ensemble) of the product of their centered states:

$$\pi_{ij} = \langle s_i s_j \rangle - \langle s_i \rangle \langle s_j \rangle \tag{D.11}$$

Note that in mathematical terms, this is the covariance of the states, but here states can only take values of -1 and +1, so this magnitude will also be bounded. The fluctuation-dissipation relation establishes the correspondence between the correlation function $\pi_{ij}$ and the magnetic susceptibility $\chi_{ij}$ with zero external field (Mora, 2007):

$$\chi_{ij} = \left.\frac{\partial m_i}{\partial h_j}\right|_{h=0} = \pi_{ij} \tag{D.12}$$

Then, the expression for $\chi_{ij}$ represents the influence of the local field of node *j* on the local magnetization value of node *i*, and is given by:

$$\chi_{ij} = \left(\beta \delta_{ij} + \sum_{l \in N(i)} \frac{\partial u_{l \to i}}{\partial h_j}\right)\left(1 - m_i^2\right) \tag{D.13}$$

For computing this local susceptibility, new messages are defined ($g_{i \to j,k}$ and $v_{i \to j,k}$), given by the derivatives of the BP messages ($h_{i \to j}$ and $u_{i \to j}$), respectively:

$$g_{i \to j,k} = \frac{\partial h_{i \to j}}{\partial h_k} \qquad\qquad v_{i \to j,k} = \frac{\partial u_{i \to j}}{\partial h_k} \tag{D.14}$$

That can be found deriving Eq. (D.7) and Eq. (D.8):

$$\boxed{g_{i \to j,k} = \beta \delta_{ik} + \sum_{l \in N(i) \setminus j} v_{l \to i,k}} \tag{D.15}$$

$$v_{j\to i,l} = \frac{\partial u_{j\to i}}{\partial h_l} = \frac{\partial}{\partial h_l}\left\{\tanh^{-1}\left[\tanh\left(\beta J_{ji}\right)\tanh\left(h_{j\to i}\right)\right]\right\} = \frac{1}{\cosh^2\left(h_{j\to i}\right)}\frac{\partial h_{j\to i}}{\partial h_l}\frac{1}{1-\tanh^2\left(\beta J_{ji}\right)\tanh^2\left(h_{j\to i}\right)}\tanh\left(\beta J_{ji}\right)$$

$$= \left(\beta\delta_{jl} + \sum_{k\in\partial j\backslash i}\frac{\partial u_{k\to j}}{\partial h_l}\right)\tanh\left(\beta J_{ji}\right)\frac{1-\tanh^2\left(h_{j\to i}\right)}{1-\tanh^2\left(u_{j\to i}\right)}$$

$$\boxed{v_{i\to j,k} = g_{i\to j,k}\tanh\left(\beta J_{ij}\right)\frac{1-\tanh^2\left(h_{i\to j}\right)}{1-\tanh^2\left(u_{i\to j}\right)}} \tag{D.16}$$

Finally, using the definition of $v_{i\to j,k}$ Eq. (D.14) in Eq. (D.13) the long-range correlations are determined from:

$$\boxed{\chi_{ij} = \left(\beta\delta_{ij} + \sum_{l\in N(i)} v_{l\to i,j}\right)\left(1-m_i^2\right)} \tag{D.17}$$

## E. Pseudo code for the SP algorithm

***ALGORITHM E.2. SUSCEPTIBILITY PROPAGATION. THE STEPS 1 TO 8 CORRESPOND TO THE BP ALGORITHM.***

| | |
|---|---|
| **STEP 1:** | Initialize all messages randomly $h_{i\to j}, u_{i\to j}, v_{i\to j,k}, g_{i\to j,k}$ |
| **STEP 2:** | **for** $i = 0$ to IMAX do // IMAX is the maximum number of iterations (**BP**) |
| **STEP 3:** | The messages corresponding to the BP given by the Eq. (D.7) and Eq. (D.8) are updated |
| **STEP 4:** | **if** convergence condition holds, **then** |
| **STEP 5:** | $i = $ IMAX // stop the iteration |
| **STEP 6:** | e**nd if** |
| **STEP 7:** | **end for** |
| **STEP 8:** | Local activations are calculated from Eq. (D.10). |
| **STEP 10:** | **for** $i = 0$ to IMAX do // IMAX is the maximum number of iterations (**SP**) |
| **STEP 11:** | The messages corresponding to the SP given by the Eq. (D.15) and Eq. (D.16) are updated. |
| **STEP 12:** | **if** convergence condition holds, **then** |
| **STEP 13:** | $i = $ IMAX // stop the iteration |
| **STEP 14:** | **end if** |
| **STEP 15:** | **end for** |
| **STEP 16:** | **end if** |
| **STEP 15:** | The correlations between all the system nodes are calculated according to Eq. (D.17) |

## F. Normalized Mutual Information

The Normalized Mutual Information (NMI) is based on the definition of a confusion matrix $\hat{\mathrm{X}}$, where the rows correspond to the real communities, and the columns to the communities found (Danon et al., 2005). The element $\mathrm{X}_{ij}$ is the number of nodes in the real community *i* that appear in the community found $j$. Finally, the equation of the measure of similarity between two partitions $A$ and $B$ is:

$$I(A,B)=\frac{-2\sum_{i=1}^{c_A}\sum_{j=1}^{c_B}\mathrm{X}_{ij}\log\left(\frac{\mathrm{X}_{ij}\mathrm{X}}{\mathrm{X}_i\mathrm{X}_j}\right)}{\sum_{i=1}^{c_A}\mathrm{X}_i\log\left(\frac{\mathrm{X}_i}{\mathrm{X}}\right)+\sum_{j=1}^{c_B}\log\left(\frac{\mathrm{X}_j}{\mathrm{X}}\right)}$$

where $c_A$ represents the number of real communities and the number of communities found is denoted by $c_B$. The sum over row *i* of matrix $[\mathrm{X}_{ij}]$ is denoted by $\mathrm{X}_i$ and the sum over column *j* by $\mathrm{X}_j$., while X denotes the Frobenius norm of the whole matrix. If the partitions found are identical to the real communities, then $I(A,B)$ takes the maximum value of 1. If the partition found by the algorithm is totally independent of the real partition, for example, when the entire network is found as a single community, $I(A,B) = 0$.

## G. Critical point of the Human Connectome and the Random Connectomes RC1 and RC2

**Fig. G.1** shows the results of the implementation of the Metropolis algorithm on the HC, the RC1 and the RC2 for computing the system susceptibility in a grid of one hundred values (from 0.01 to 1, with step of 0.01) of the temperature (inverse of *β*). The critical parameters are similar ($T_C = 0.40$ for HC, $T_C = 0.42$ for RC1 and $T_C = 0.43$ for RC2), since the networks share topological properties and similarities in the values of the weights.

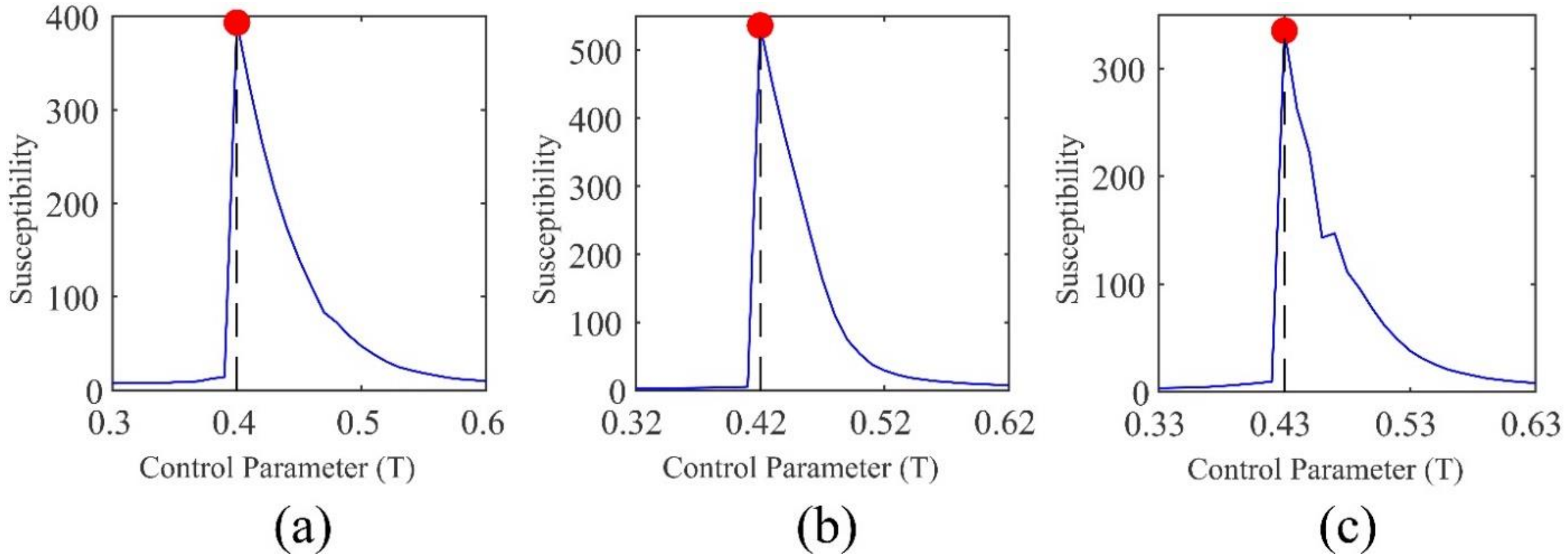

**Fig. G.1.** The three graphs show the curve of the system susceptibility ($\chi$) fir different values of T, obtained by the Metropolis algorithm for: a) the Human Connectome, b) the Random Connectome 1 and c) the Random Connectome 2. The abscissa value of the red dot represents the critical parameter of the corresponding network: $T_C = 0.40$ for HC, $T_C = 0.42$ for RC1 and $T_C = 0.43$ for RC2.

## H. Comparing clustering of the HC and of the functional connectivity obtained with SP from the null-model network RC2, with empirical resting-state networks (RSNs)

**Fig. H1-H2** show -as brain activation maps- the modules of the connectivity matrix obtained with the SP on the RC2 and the modules obtained directly from the HC anatomical matrix, respectively. The modules were matched to the experimental RSNs according to the highest overlapping between them. The experimental activations of the RSNs were represented with the masks described in (Beckmann et al., 2005), which were translated to the Human Connectome of 998 regions in (Haimovici et al., 2013). These two figures allow to confirm visually that the SP-HC (**Fig. 8**) is the one with better overall match.

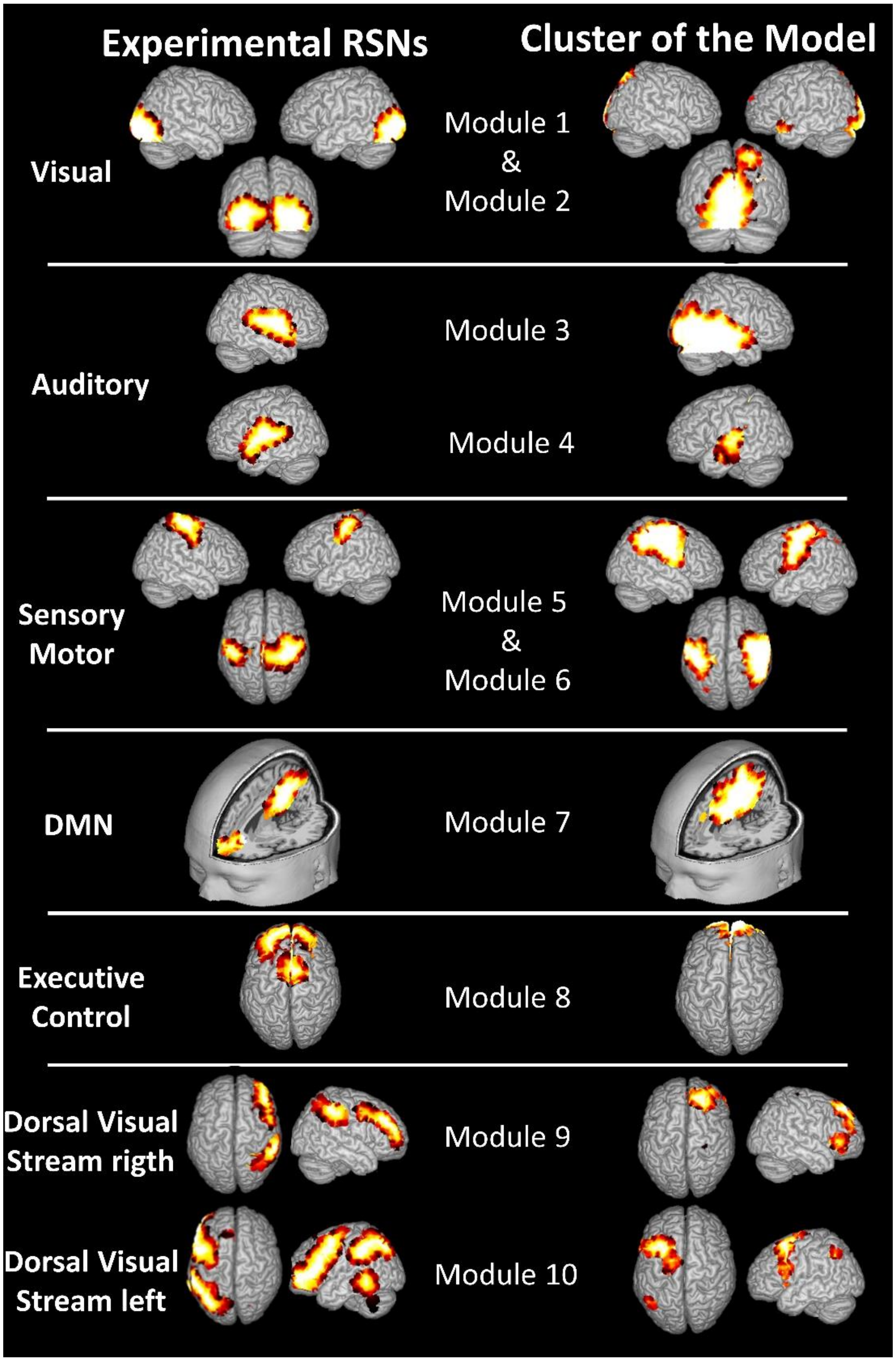

**Fig. H.1.** Modules derived from the correlation matrix (right column) obtained with SP on the null hypothesis network RC2 (SP-RC2, **Fig. 5c**) and the spatial maps of experimental functional resting-state networks (left column).

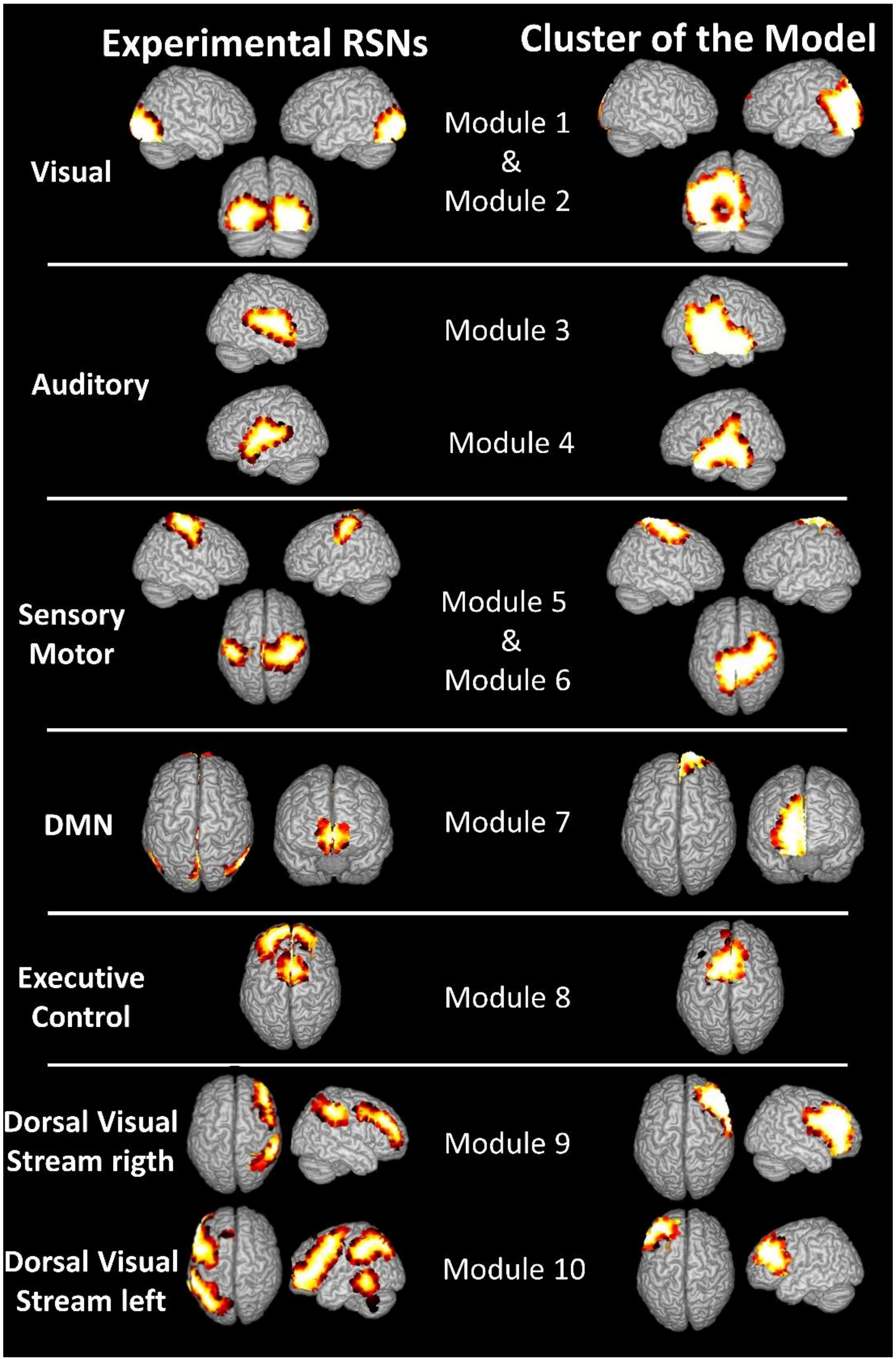

**Fig. H.2.** Modules derived from the correlation matrix (right column) obtained with SP on the Human Connectome (HC, **Fig. 5d**) and the spatial maps of experimental functional resting-state networks (left column).

## I. Modeling resting-state activity and connectivity in two pathologies

The simulation of a structural network corresponding to a subject with Alzheimer's Disease (AD) consisted in decreasing all connection weights of the original Human connectome (HC) to their 80%. This means that the topology is exactly the same of the HC. In the case of the Callosotomy or agenesis of Corpus Callosum (CC), all connections between pairs of nodes located in different brain hemispheres were set to 0, simulating the absence of those fibers connecting both hemispheres. The topology of the network is disrupted although it is the same as the HC for the other connections, keeping even the same connection weights. **Fig. I.1** shows the connectivity matrices for these structural networks.

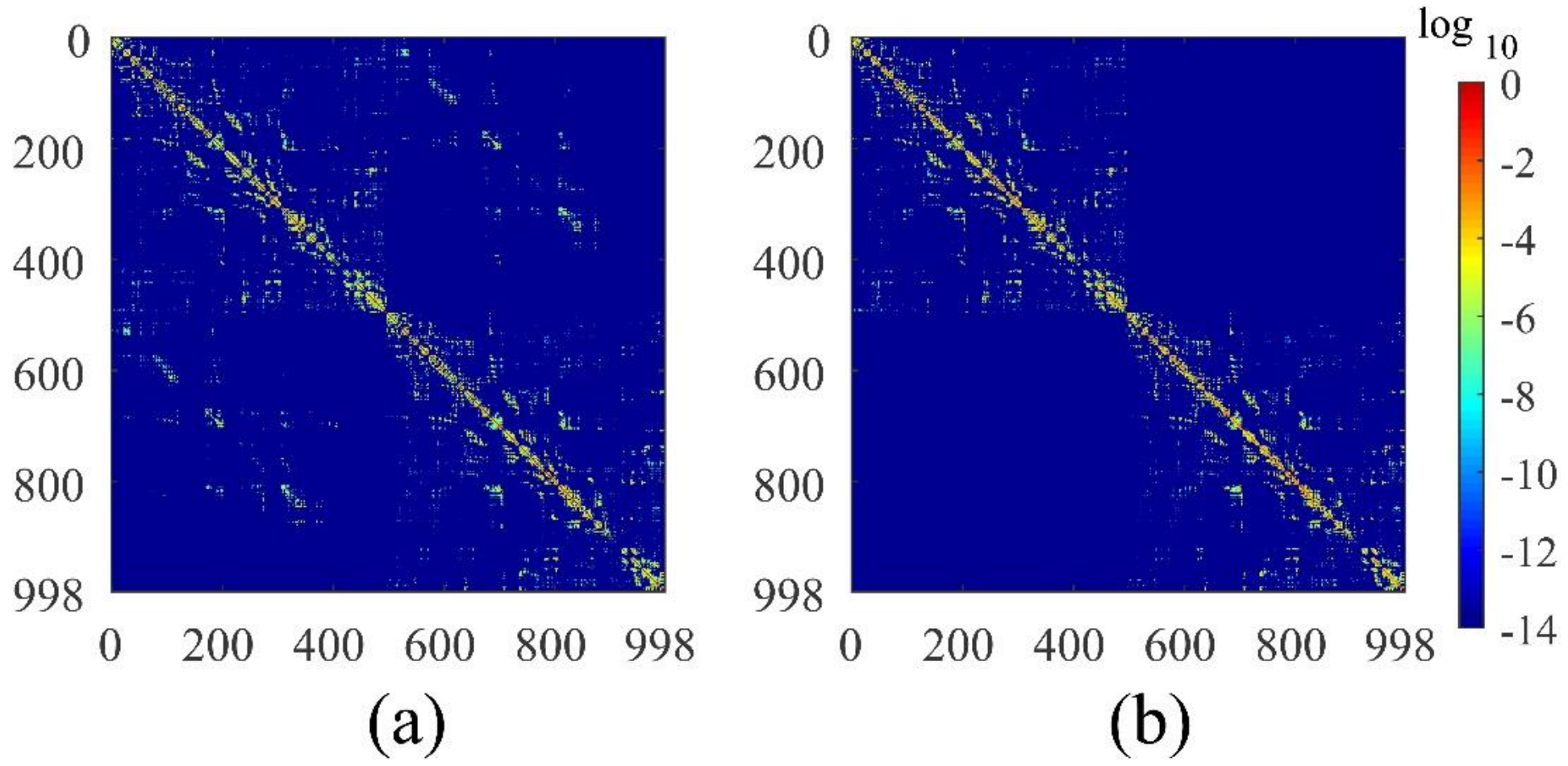

**Fig. I.1.** Anatomical connectivity matrices created for simulating structural networks of Alzheimer's Disease (a) and the absence of the Corpus Callosum (b).

For computing the functional maps as the fixed-point solution of BP and the functional connectivity matrices with SP, we decided to use the critical control parameter of the HC, in order to simulate the immediate functional alterations that may appear from disruptions of the anatomical network. However, it is also interesting to confirm what happen in the case of using the critical value of the control parameter found for these altered structural networks. **Fig. I.2** shows the functional maps and **Fig. I.3** shows the modules obtained from the connectivity matrices estimated with the BP and SP respectively, for the two pathological cases. As expected, using the critical control of the AD altered network just re-scale the connections to their critical value so, the BP and SP solutions are exactly the same as in the original HC case. In the case of CC it is more interesting to see that despite the topology of the network can never be recovered, the functional map obtained using the own critical value is much more similar to experimental Default Mode Network maps than the map obtained keeping the critical value of the HC. The corresponding modules of the connectivity matrices found with SP are also more similar to the modules of a healthy person, despite it is obvious that there are no clusters sharing nodes from different hemispheres. This result could provide an explanation, from the information theory point of view, of the functional re-organization of the brain activity after long periods of acquiring or suffering changes in the anatomical connections.

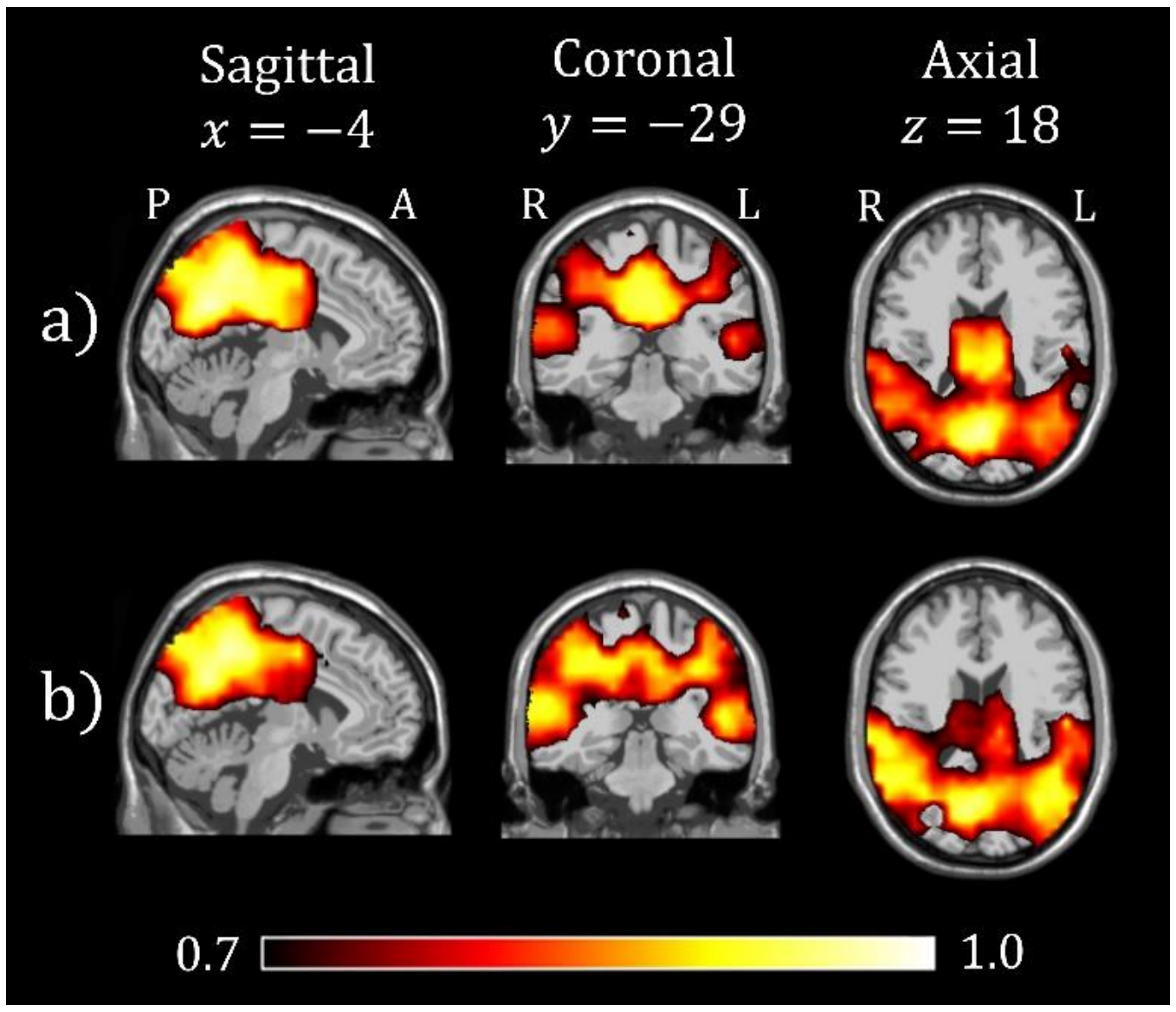

**Fig. I.2.** Functional maps obtained with the BP algorithm when using the critical value for the control parameter computed for the anatomical networks simulating Alzheimer's Disease (a) and the absence of the Corpus Callosum (b).

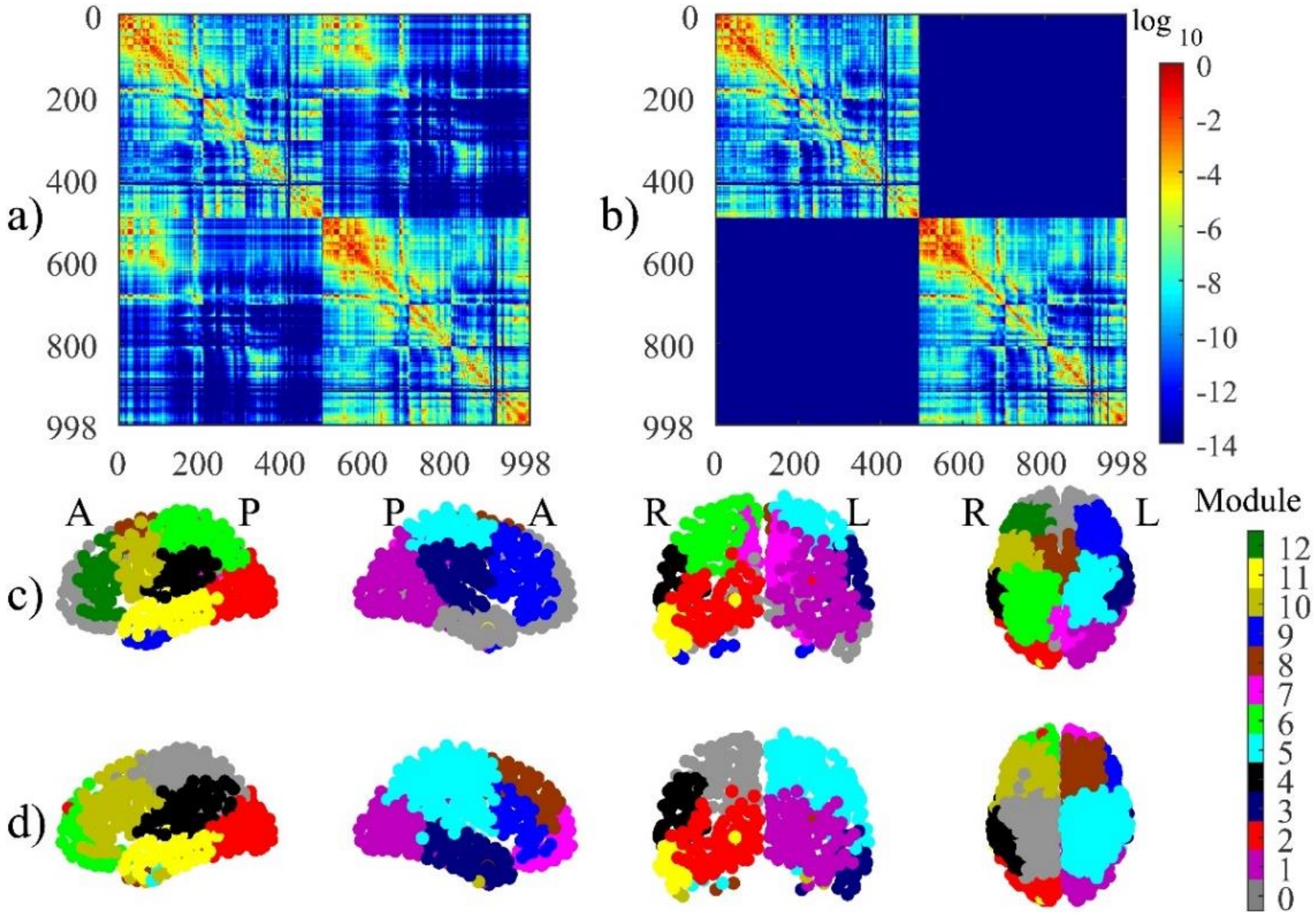

**Fig I.3.** Modularity of the connectivity matrices obtained with the SP algorithm when using the critical value for the control parameter computed for the anatomical networks simulating Alzheimer's Disease (a and c) and the absence of the Corpus Callosum (b and d).

## J. Glossary and mathematical notation

| Symbol | Description |
|---|---|
| $T$ | Control parameter |
| $T_C$ | Critical parameter |
| $N$ | Number of regions in the network |
| $s_i$ | Variable that represents the state of the brain region (node) *i* |
| $\mathcal{H}$ | Hamiltonian of the system |
| $J_{ij}$ | Interchange interaction between the brain regions i and *j* |
| $h_i$ | External field on the *i*-th brain region |
| $Z$ | Statistical sum or partition function |
| $\beta$ | Inverse of the control parameter |
| $p_i$ | Marginal distributions of probability. |
| $b_i$ | Belief of node $i$, which represents the probability that node *i* is in the state $s_i$ |
| $m_{j\to i} = m_{ji}$ | Variable that represents the message (information) sent by node *j* about which state is more favorable for node *i* |
| $N(j)$ | Represents the set of neighbors of node j |
| $k$ | normalization constant |
| $t_d$ | Refers to the discrete time system. |
| $\Delta d$ | Difference between the new and the old message after the updating. |
| $\delta_{max}$ | Convergence condition |
| $m_i$ | Magnetization of the spin *i* |
| $v_i$ | Activity of the neuron *i* in the Hopfield model |
| $f(x)$ | Activation function in the neurodynamic equation of the Hopfield model |
| $w_{ij}$ | Connection weight (synaptic) for the Hopfield model |
| $K_i(t)$ | external signal in the Hopfield model |
| $n_{ji}$ | New messages defined to establish the connection between the BP equation and the Hopfield model |
| $v_j^i$ | Electrical activity of the neuron *j* on the *i* |
| $q_{j\to i}$ | Corresponds with $m_{j\to i}$ |
| $p_{j\to i}$ | Factor corresponding to the external field in the equation for $q_{j\to i}$ |
| $h_{i\to j}$ and $u_{i\to j}$ | Re-definition of the messages $m_{i\to j}$ in the logarithmic notation of likelihood. |
| $g_{i\to j,k}$ and $v_{i\to j,k}$ | New messages defined for the SP algorithm |
| $\chi_{ij}$ | Correlation between node *i* and *j* |
| $\hat{X}$ | Confusion matrix |